\documentclass[trackchanges]{aastex6}
\usepackage{amsmath}
\usepackage{xspace}
\usepackage{longtable}
\usepackage{floatrow} 
\usepackage[colorinlistoftodos]{todonotes}
\usepackage{xcolor} 

\newcommand {\chandra} {\textsl{Chandra}}
\newcommand {\swift} {\textsl{Swift}}

\newcommand {\maxi} {\textsl{MAXI}}

\newcommand {\nustar} {\textsl{NuSTAR}}
\newcommand {\nicer} {\textsl{NICER}\xspace}
\newcommand {\atca} {\textsl{ATCA}}
\newcommand{\cxo}{{\it Chandra}\xspace} 

\newcommand{\irsf}{{\it IRSF}\xspace}
\newcommand{\gemini}{{\it Gemini}\xspace}
\newcommand{\integral}{{\it INTEGRAL}\xspace}

\newcommand{\degrees}{\mbox{$^\circ$}}
\newcommand{\E}[1]{\ensuremath{\times 10^{#1}}}
\def\xsrc{J1621}

\def\maxinos{J1621}

\begin{document}


\title{Discovery and Identification of MAXI J$1621-501$ as a type I X-ray Burster with a Super-Orbital Period}


\author{Nicholas M. Gorgone\altaffilmark{1,2}}
\author{Chryssa Kouveliotou\altaffilmark{1,2}}
\author{Hitoshi Negoro\altaffilmark{3}}
\author{Ralph A. M. J. Wijers\altaffilmark{4}}
\author{Enrico Bozzo\altaffilmark{5}}
\author{Sylvain Guiriec\altaffilmark{1,2}}
\author{Peter Bult\altaffilmark{6}}
\author{Daniela Huppenkothen\altaffilmark{7}}
\author{Ersin G\"{o}\u{g}\"{u}\c{s}\altaffilmark{8}}
\author{Arash Bahramian\altaffilmark{9}}
\author{Jamie Kennea\altaffilmark{10}}
\author{Justin D. Linford\altaffilmark{11, 12}}
\author{James Miller-Jones\altaffilmark{9}}

\author{Matthew G. Baring\altaffilmark{13}}
\author{Paz Beniamini\altaffilmark{1,2}}
\author{Deepto Chakrabarty\altaffilmark{15}}
\author{Jonathan Granot\altaffilmark{1,16}}
\author{Charles Hailey\altaffilmark{17}}
\author{Fiona A. Harrison\altaffilmark{14}}
\author{Dieter H. Hartmann\altaffilmark{18}}
\author{Wataru Iwakiri\altaffilmark{19}}
\author{Lex Kaper\altaffilmark{4}}
\author{Erin Kara\altaffilmark{20}}
\author{Simona Mazzola\altaffilmark{21}}
\author{Katsuhiro Murata\altaffilmark{22}}
\author{Daniel Stern\altaffilmark{23}}
\author{John A. Tomsick\altaffilmark{24}}
\author{Alexander J. van der Horst\altaffilmark{1,2}}
\author{George A. Younes\altaffilmark{1,2}}

\altaffiltext{1}{Department of Physics, The George Washington University, Washington, DC 20052, USA}
\altaffiltext{2}{Astronomy, Physics and Statistics Institute of Sciences (APSIS), The George Washington University, Washington, DC 20052, USA}
\altaffiltext{3}{Department of Physics, Nihon University, 1-8 Kanda-Surugadai, Chiyoda-ku, Tokyo 101-8308, Japan}
\altaffiltext{4}{University of Amsterdam, Science Park 904, 1098 XH Amsterdam, The Netherlands}
\altaffiltext{5}{University of Geneva, Chemin d' Ecogia 16, Versoix, 1290 Switzerland}
\altaffiltext{6}{Astrophysics Science Division, NASA’s Goddard Space Flight Center, Greenbelt, MD 20771, USA}
\altaffiltext{7}{DIRAC Institute, Department of Astronomy, University of Washington, 3910 15th Ave NE, Seattle, WA 98195}
\altaffiltext{8}{Faculty of Engineering and Natural Sciences, Sabanc\i{} University, Orhanl\i-Tuzla 34956, \.{I}stanbul, Turkey}
\altaffiltext{9}{International Centre for Radio Astronomy Research -- Curtin University, GPO Box U1987, Perth, WA 6845, Australia}
\altaffiltext{10}{The Pennsylvania State University, University Park, PA 16802, USA}
\altaffiltext{11}{Department of Physics and Astronomy, West Virginia University, P.O. Box 6315, Morgantown, WV 26506, USA}
\altaffiltext{12}{Center for Gravitational Waves and Cosmology, West Virginia University, Chestnut Ridge Research Building, Morgantown, WV 26505, USA}
\altaffiltext{13}{Physics and Astronomy Department, Rice University, 6100 Main Street, Houston, TX, USA}
\altaffiltext{14}{Cahill Center for Astrophysics, California Institute of Technology, 1216 East California Boulevard, Pasadena, CA 91125, USA}
\altaffiltext{15}{MIT Kavli Institute for Astrophysics and Space Research, Massachusetts Institute of Technology, Cambridge, MA 02139, USA}
\altaffiltext{16}{Department of Natural Sciences, The Open University of Israel, P.O Box 808, Ra'anana 43537, Israel}
\altaffiltext{17}{Columbia Astrophysics Laboratory, Columbia University, New York, NY 10027, USA}
\altaffiltext{18}{Department of Physics and Astronomy, Clemson University, Kinard Lab of Physics, Clemson, SC 29634-0978, USA}
\altaffiltext{19}{Department of Physics, Faculty of Science and Engineering, Chuo University, 1-13-27 Kasuga, Bunkyo-ku, Tokyo 112-8551, Japan}
\altaffiltext{20}{Department of Astronomy, University of Maryland, College Park, MD 20742}
\altaffiltext{21}{Dipartimento di Fisica e Chimica - Emilio Segr\`{e}, Universit\`{a} degli Studi di Palermo, Via  Archirafi 36 - 90123 Palermo, Italy}
\altaffiltext{22}{Department of Particle and Astrophysical Science, Nagoya University, Furo-cho, Chikusa-ku, Nagoya, 464-8602, Aichi, Japan}
\altaffiltext{23}{Jet Propulsion Laboratory, California Institute of Technology, 4800 Oak Grove Drive, Mail Stop 169-221, Pasadena, CA 91109, USA}
\altaffiltext{24}{Space Sciences Laboratory, 7 Gauss Way, University of California, Berkeley, CA 94720-7450, USA}


\begin{abstract}
MAXI J1621--501 is the first \swift/XRT Deep Galactic Plane Survey transient that was followed up with a multitude of space missions (\nustar{}, \swift{}, \chandra{}, \nicer{}, \integral{}, and \maxi{}) and ground-based observatories (\gemini, \irsf, and \atca). The source was discovered with \maxi{} on 2017 October 19 as a new, unidentified transient. Further observations with \nustar{} revealed 2 Type I X-ray bursts, identifying MAXI J1621-501 as a Low Mass X-ray Binary (LMXB) with a neutron star primary. Overall, 24 Type I bursts were detected from the source during a 15 month period. At energies below 10 keV, the source spectrum was best fit with three components: an absorbed blackbody with $kT=2.3$\,keV, a cutoff power law with index $\Gamma=0.7$, and an emission line centered on $6.3$\,keV. Timing analysis of the X-ray persistent emission and burst data has not revealed coherent pulsations from the source or an orbital period. We identified, however, a super-orbital period $\sim78$\, days in the source X-ray light curve. This period agrees very well with the theoretically predicted radiative precession period of $\sim82$\,days. Thus, MAXI J1621-501 joins a small group of sources characterized with super-orbital periods.

\end{abstract}


\section{Introduction}
Low mass X-ray binaries (LMXBs) consist of a compact object, either a neutron star (NS) or a black hole (BH), and a donor star, typically a late-type, low-mass main sequence star. Most LMXBs are persistent X-ray sources, as mass transferred through Roche lobe overflow from the donor to the compact object efficiently converts its gravitational potential energy to X-rays \citep{Remillard+McClintock2006, Done+2007}. Several LMXBs exhibit transient outbursts, during which their X-ray luminosities increase by orders of magnitude; such outbursts generally last anywhere between a couple of weeks to months. Identifying the nature of the compact object, namely whether it is a NS or a BH, through data collected during the outburst, is usually a non-trivial task. However, the detection of type I X-ray bursts from the compact object, unambiguously identifies it as a NS. Type I X-ray bursts are thermonuclear explosions; they take place on the surface layers of the NS when the accreted He or H from the companion star reaches the critical density to initiate nuclear fusion, resulting in a He flash.

About $\sim 30$ transient LMXBs have exhibited multi-episodic rebrightening during their outbursts that arises on long timescales known as ``super-orbital periods" or long periods. These are not strictly periodic and they are likely due to a long period present in the system \citep{Gerend+Boynton1976} produced by the accretion disk when a torus of disk material precesses retrogradely. The warped disk can reflect or obscure light from the accretor, depending on its orientation (see Figs. 7 and 10 of \citet{Wijers+Pringle1999}, hereafter WP99). \maxinos{} exhibits such a long period, which we discuss in \autoref{discussion}.
		


On 2017 October 19, the Monitor of All-sky X-ray Image (\maxi{}) Nova Alert System \citep{discoveryATEL, maxi_nova_alert_system} was activated by a new transient designated as MAXI J1621--501 (hereafter J1621). This observation was followed up with a Neil Gehrels \swift{} Observatory (hereafter \swift{}) Target of opportunity (ToO) observation to localize the source, which fell within the survey area of the \swift{}/XRT Deep Galactic Plane Survey (DGPS). The DGPS is a \swift{} and \nustar{} legacy project designed to systematically search for transients within the Galactic boundaries of $\lvert{}b{}\rvert{}<0.5^{\circ{}}$, $10^{\circ{}}<\lvert{}\ell{}\rvert{}<30^{\circ{}}$ (initial phase, PI: C. Kouveliotou; \citealt{DGPS}). \maxinos{} is the first new source classified with the DGPS.

The \swift{} and \integral{} teams initiated monitoring campaigns of \maxinos{}, soon after its discovery, which were interspersed with multi-wavelength ToO requests. X-ray observations were interrupted for a 2.5 month-long period (late October 2017 to mid January 2018) due to Sun constraints. \maxi{} and \swift/XRT resumed monitoring thereafter; thus far, \maxinos{} has exhibited a series of six weaker outbursts (Fig. \ref{fig: maxi lightcurve}).
The \maxi\ data show that the source activity subsided as of March 2019, returning to its quiescent count rate $\sim{}10^{-2}$~cts s$^{-1}$.

In early December 2017, we activated the \nustar{} legacy program to observe \maxinos. These observations revealed two type I X-ray bursts, enabling the concrete identification of \maxinos{} as a type I X-ray burster. Further observations with \integral, \maxi{} and the Neutron Star Interior Composition Explorer (\nicer), detected a total of 24 bursts. In early March 2018 we observed and accurately localized the source with our \cxo ToO observation. The \cxo location enabled the solid identification of a near-IR counterpart in our \gemini{} follow up observations, which was clearly brighter than its likely quiescent state as observed in archival data (Bahramian et al., in preparation).


We describe below our comprehensive, 15-month long X-ray campaign monitoring the outburst of \maxinos{} as well as our multi-wavelength searches during this interval. In \autoref{instrumentation} we discuss the observations and the data processing, and in \autoref{analysis} we describe the spectral and temporal results of the source persistent emission. We present detailed analysis of 3 out of the 24 type I X-ray bursts \citep{bult_atel_burst, integral_sees_burst} in \autoref{sec: bursts}. We discuss our results in \autoref{discussion}.  


\section{Observations and Data Processing} \label{instrumentation} 

We observed \maxinos{} with multiple X-ray missions \nustar{}, \swift{}, \chandra{}, \nicer{}, \integral{}, and \maxi{} to trace the X-ray temporal and spectral evolution throughout its outburst. For the localization of the source  X-ray counterpart we employed our ToO observation with \chandra{}/HRC-I, which led to confirmation of its near-IR counterpart with \gemini{}. We also utilized \swift{}/UVOT and Infra-Red Survey Facility (\irsf{}) observations, as well as observations at longer wavelengths with the Australia Telescope Compact Array (\atca{}). We describe below our observations and data processing for each instrument. Table \ref{tbl: timeline} lists the timeline of all observations per instrument.

\begin{table}[!t]
\begin{tabular}{lclccrc}
Obs. & ID & Mission & Telescope/Mode & Start Time [UT] & Exposure & Reference \\ 
     &    &         &                & [dd Mmm yyyy hh:mm] & [ks]   &           \\ \hline\hline
1.$\dagger{}$ & - & \maxi{} & GSC & 19 Oct. 2017 05:45   & * & [1] \\
2.  & 00010351007 & \swift{} & XRT/PC + UVOT & 19 Oct. 2017 18:09   & 0.1 & [2]\\
3.  & 00010352001 & \swift{} & XRT/PC + UVOT & 19 Oct. 2017 18:11   & 0.4 & [2]\\
4. & 00010357001 & \swift{} & XRT/PC + UVOT & 19 Oct. 2017 19:17   & 0.5 & [2]\\
5. & 00036140002 & \swift{} & XRT/WT + UVOT & 20 Oct. 2017 20:39   & 1.0 & This work \\
6. & 1020630101 & \nicer{}  &  XTI & 21 Oct. 2017 14:24   & 0.1 & This work \\
7. & GS-2018A-DD-201 & \gemini{} & FLAMINGOS-2 & 21 Oct. 2017 23:37   & $\ddagger{}$ & [3] \\
8. & 00036140003 & \swift{} & XRT/WT & 22 Oct. 2017 01:56   & 0.4 & This work \\
9. & 1020630102 & \nicer{}  &  XTI & 22 Oct. 2017 09:25   & 0.6 & This work \\
10. & - & \irsf{} & SIRIUS & 22 Oct. 2017 17:44 & 0.3 & This work \\
11. & 00036140004 & \swift{} & XRT/WT + UVOT & 24 Oct. 2017 04:35   & 1.1 & This work \\
12. & 90301322001 & \nustar{} & FPMA/B & 26 Oct. 2017 00:46   & 38.3 & This work \\
13. & 00087355002 & \swift{} & XRT/PC + UVOT & 26 Oct. 2017 07:25   & 0.6 & [4]\\
14. & CX399 & \atca{} & 5.5, 9 GHz & 18 Nov. 2017 19:00   & 18.0 & This work \\
15.$\dagger{}$ & 9030132800[1/2] & \nustar{} & FPMA/B & 02 Dec. 2017 23:06   & 39.0 & [5] \\
16. & 9030132800[3/4] & \nustar{} & FPMA/B & 03 Dec. 2017 15:16   & 10.0 & This work \\
17. & 00036140006 & \swift{} & XRT/WT + UVOT & 13 Jan. 2018 11:47   & 0.9 & This work \\
18. & 00036140007 & \swift{} & XRT/WT + UVOT & 20 Jan. 2018 19:05   & 0.9 & This work \\
19. & 00036140008 & \swift{} & XRT/WT + UVOT & 27 Jan. 2018 00:51   & 1.0 & This work \\
20. $ \dagger{}$  & -        & \integral{} & ISGRI/JEM-X(1 \& 2) & 27 Jan. 2018 14:29 & * & [6] \\
21. & 00036140009 & \swift{} & XRT/WT & 03 Feb. 2018 03:44   & 1.0 & This work \\
22. & 00036140011 & \swift{} & XRT/PC + UVOT & 08 Feb. 2018 12:29   & 1.0 & [7]\\
23. & 00036140012 & \swift{} & XRT/WT + UVOT & 10 Feb. 2018 15:41   & 0.8 & This work \\
24. & 00036140014 & \swift{} & XRT/WT + UVOT & 21 Feb. 2018 05:57   & 0.6 & This work \\
25. & 20334 & \chandra{} & HRC-I & 22 Feb. 2018 20:43   & 2.6 & This work \\
26. $\dagger{}$ & 1034170101 & \nicer{}  &  XTI & 22 Feb. 2018 16:04   & 5.7 & This work \\
27. & 1034170102 & \nicer{}  &  XTI & 22 Feb. 2018 23:59   & 0.5 & This work \\
28. & 00036140016 & \swift{} & XRT/WT + UVOT & 28 Feb. 2018 09:16   & 0.9 & This work \\
29. & 1034170103 & \nicer{}  &  XTI & 28 Feb. 2018 07:49   & 4.2 & This work \\
30. & 1034170104 & \nicer{}  &  XTI & 01 Mar. 2018 06:58   & 5.3 & This work \\
31. & 1034170105 & \nicer{}  &  XTI & 02 Mar. 2018 00:02   & 5.7 & This work \\
32. & 1034170106 & \nicer{}  &  XTI & 03 Mar. 2018 16:01   & 2.4 & This work \\
33. & 1034170107 & \nicer{}  &  XTI & 04 Mar. 2018 10:32   & 2.1 & This work \\
34. & 1034170108 & \nicer{}  &  XTI & 05 Mar. 2018 06:43   & 1.7 & This work \\
35. & 1034170109 & \nicer{}  &  XTI & 06 Mar. 2018 15:08   & 1.3 & This work \\
36. & 00036140017 & \swift{} & XRT/WT + UVOT & 08 Mar. 2018 21:15   & 0.8 & This work \\
37. & 00036140018 & \swift{} & XRT/WT + UVOT & 10 Mar. 2018 22:39   & 1.1 & This work \\
38. & 00010670001 & \swift{} & XRT/WT + UVOT & 25 Apr. 2018 10:42   & 1.5 & This work \\
39. & \hspace{0.3cm} 00010670002 \hspace{0.3cm} & \swift{} \hspace{1cm}  & XRT/WT + UVOT & 27 Apr. 2018 12:09   & 1.1 & - \\
40. & 00010670003 & \swift{} & XRT/WT + UVOT & 02 May 2018 22:39  & 0.9 & This work \\
    &             &          & $\vdots{}$    &                      &     &   \\

\end{tabular}
\caption{Observations of MAXI J1621--501 in chronological order}
\tablecomments{*~continuous monitoring, $\dagger{}$~X-ray bursts detected, $\ddagger{}$~Exposures are 0.3~ks, 0.1~ks, and 0.3~ks for the H, J, and K$_{s}$ bands, respectively.}
\tablerefs{[1] \citet{discoveryATEL}, [2] \citet{swift_tile_ATEL}, [3] Bahramian et al. (in prep), [4] \citet{swift_as_DGPS_ATEL}, [5] \citet{bult_atel_burst}, [6] \citet{integral_sees_burst}, [7] \citet{INTEGRAL_disambig_atel}}
\label{tbl: timeline}
\end{table}

\begin{table}[!t]
\begin{tabular}{lclccrc}
Obs. & ID & Mission & Telescope/Mode & Start Time [UT] & Exposure & Reference \\ 
     &    &         &                & [dd Mmm yyyy hh:mm] & [ks]   &           \\ \hline\hline   
41. & 00010670004 & \swift{} & XRT/WT + UVOT & 09 May 2018 12:37 & 0.5 & This work \\
42. & 00010670005 & \swift{} & XRT/WT + UVOT & 17 May 2018 08:33 & 0.9 & This work \\
43. & 00010670006 & \swift{} & XRT/WT + UVOT & 23 May 2018 06:26 & 1.0 & This work \\
44. & 00010670007 & \swift{} & XRT/WT + UVOT & 30 May 2018 16:47 & 0.9 & This work \\
45. & 1034170110 & \nicer{}  &  XTI & 02 Jun. 2018 23:10   & 0.4 & This work \\
46. & 1034170111 & \nicer{}  &  XTI & 03 Jun. 2018 03:48   & 0.2 & This work \\
47. & 1034170112 & \nicer{}  &  XTI & 05 Jun. 2018 14:14   & 1.2 & This work \\
48. & 00010670008 & \swift{} & XRT/WT + UVOT & 06 Jun. 2018 10:02 & 1.0 & This work \\
49. & 00010670009 & \swift{} & XRT/WT + UVOT & 13 Jun. 2018 04:52 & 0.03 & This work \\
50. & 00010670010 & \swift{} & XRT/WT + UVOT & 18 Jun. 2018 10:52 & 0.8 & This work \\ 
51. & 00010670011 & \swift{} & XRT/WT + UVOT & 20 Jun. 2018 05:43 & 0.3 & This work \\ \hline\hline
\end{tabular}
\end{table}


\subsection{The Monitor of All-sky X-ray Image (\maxi{})}
\maxi{} \citep{maxi} is an all-sky monitor mounted on the Japanese Experimental Module Exposed Facility of the International Space Station (ISS). \maxi{} covers 85\% of the sky at a cadence of $\sim{}$\hspace{-1mm}90 min. The monitor consists of two instruments, the Solid-State Slit Camera (SSC: $0.7-7$~keV; \citealt{maxi_ssc}) and the Gas Slit Camera (GSC: $2-20$~keV; \citealt{maxi_gsc}). 

To analyze the observations of \maxinos{} we first extracted \maxi{}/GSC data in the 2--4 keV (soft) and 4--10 keV (hard) bands by using the image fit method \citep{Morii2017}, which takes into account the point spread functions of the cameras, as well as X-ray contamination from nearby sources. Due to 4U 1624$-$490 lying 1.49\degrees{} away, we only used cameras GSC\_2, GSC\_4, GSC\_5 and GSC\_7, which are well-calibrated spatially. We then subtracted the Galactic ridge emission, comprising constant contributions of about 3.2\E{-3}~cts~s$^{-1}$~cm$^{-2}$ and 5.2\E{-3}~cts~s$^{-1}$~cm$^{-2}$ in the soft and hard energy band, respectively, and sinusoidal components with amplitudes of 4.0\E{-3} cts~s$^{-1}$~cm$^{-2}$ and 3.5\E{-3} cts~s$^{-1}$~cm$^{-2}$, respectively. The latter component has a period of 72.14 days, possibly resulting from the ISS orbital precession. All background contributions were estimated from MJD 57000 to MJD 57999. For the $2-4$ keV light curve, we added a systematic uncertainty of 10\%, obtained through the same image fit analysis of the Crab Nebula. Some data points were unusable due to image-fit results affected by flux variations of the nearby source 4U 1624$-$490. Fig. \ref{fig: maxi lightcurve} shows the count rate evolution of \maxinos{} in all three energy ranges (hard, soft, and total). 

\subsection{The Neil Gehrels Swift Observatory/X-ray Telescope (\swift/XRT)}
\swift{}/XRT \citep{swiftXRT} was used in photon counting (PC) and window timing (WT) modes. For our spectral analysis we corrected for pileup in the PC mode by pairing annular extraction regions around the source with ancillary response files\footnote{See the \href{http://www.swift.ac.uk/analysis/xrt/arfs.php}{\swift{} Leicester Site} for details (http://www.swift.ac.uk/analysis/xrt/arfs.php)}. No WT mode count rates exceeded the 100 cts s$^{-1}$ threshold for pileup \citep{Romano2006}. 

The \swift{} observations provide sporadic coverage of the \maxinos{} outburst. They are distributed in three intervals: the very beginning of the outburst (October 2017), mid-January to mid-February 2018, and March to June 2018. We flag  Obs. 2, 11, and 35  in Table \ref{tbl: timeline} as unusable for scientific analysis. In obs. 2 we were not able to extract photons in a annulus of sufficient width for a spectral analysis or flux estimation, due to the source off-axis location ($\sim{}$11.3\arcmin{}), the short exposure time (0.1ks), and the significant pileup ($\sim{}$3.7~cts s$^{-1}$). Obs. 11 was not used as it was on the edge of the 1D field of view, and obs. 35 was not used due to a short exposure time of 42~s.

To fit the \swift{}/XRT PC mode data, we extracted a centroided, circular source region of radius=20~pixels  $\sim{}50^{\prime{}\prime{}}$ and a  circular background source region, at a similar off-axis angle, with radius of at least 50$^{\prime{}\prime{}}$, depending on the actual source position within the field of view. We then extracted spectra and lightcurves with \texttt{xselect} for source and background, created an exposure map, and created ancillary response files for source and background. Using \texttt{grppha} we grouped the spectra at a minimum of 10 to 30 photons per bin, depending on the total source photon count.

To fit the \swift{}/XRT WT mode data, we carried out a similar process, using a centroided, rectangular source region of length 20~pixels$\sim{}$50$^{\prime{}\prime{}}$ and a rectangular background region with length of at least 50$^{\prime{}\prime{}}$, depending on the source distance from the central pixel of the 1D projection. We used the same methods as for the PC-mode data to extract and prepare the spectra for fitting.

\subsection{The Nuclear Spectroscopic Telescope ARray (\nustar{})} \label{subsec:Nustar_descrip}
We observed \maxinos{} with \nustar{} \citep{nustar} three times for a total exposure of $\sim$90~ks. \nustar{} consists of two co-aligned, grazing-incidence Wolter-I Focal Plane Modules (FPMA/B). Here we flag \nustar{} obs. 15 and 16, which were carried out in data mode 06, while J1621 was $28^{\circ{}}$ from the Sun. In this mode, positional information is only accurate to 2\arcmin{}, instead of the nominal 8\arcsec{}. Each observation was divided into Good Time Intervals (GTIs), where the aspect solution was determined by different combinations of Camera Header Units (CHUs)\footnote{See \S{}6.7 of the \href{https://heasarc.gsfc.nasa.gov/docs/nustar/analysis/nustar_swguide.pdf}{NuSTAR Data analysis Software Guide} (https://heasarc.gsfc.nasa.gov/docs/nustar/analysis/nustar\_swguide.pdf)}. 

We extracted the data for \nustar{} obs. 15 and 16 by first splitting the cleaned Level 2 event file using the \textsc{nusplitsc} command, which produced one event file for each CHU combination (CHUs 2, 3, 1+2, 1+3, and 2+3 were used). For each event file, we created with \textsc{ds9} circular source regions centered on \maxinos{} with radius=120$^{\prime{}\prime{}}$. This process allowed for visually smooth transitions between CHU switches in the source persistent emission lightcurve and usually allows $\sim{}85\%$ enclosed energy fraction \citep{An2014}. It is expected that the data mode 06 encloses a lower percentage of the overall point source energy. A background region file of the same shape and size was created near the source. We then ran the standard \textsc{nuproducts} command from \textsc{heasoft v}6.22 on each event file to extract lightcurves and spectra. Spectral fits to \nustar{} data were limited due to high background above 25~keV during our observations.

To extract the spectra in the first \nustar{} epoch (Obs. 12, Tbl. \ref{tbl: timeline}) we created a circular source region centered on the \nustar{} centroid with radius (r=120\arcsec{}). A background region of the same size and shape was constructed. A response matrix and ancillary response file was created for each spectrum using the \textsc{nuproducts} routine in \textsc{HEASOFT}. For the other two \nustar{} epochs (Obs 15 and 16, Tbl. \ref{tbl: timeline}), which were taken in mode 06, we extracted spectra separately for each of the 5 CHU combinations for both focal plane modules A and B. All ten were fit together, totaling 20 spectra. To avoid mixing the spectra, an ancillary response file and response matrix file were generated for each spectrum individually. Spectrum extraction was done with the same source and background shapes and sizes, but the source centers were chosen to be the \nustar{} data mode 06 source centroid for each combination of CHUs.

\subsection{The Neutron Star Interior Composition Explorer (\nicer{})}
\label{sec:nicer data}

Also mounted on the ISS, \nicer{} \citep{nicer} comprises 56 co-aligned X-ray concentrator optics, each paired with a single pixel silicon drift detector sensitive in the $0.2-12$ keV passband \citep{Prigozhin2012}. We started observing \maxinos{} on 2017 October 21, however, due to limited source visibility, only about 700~s of exposure could be collected at that time. Additional observations were collected in 2018 February, March and June. The \nicer data are available under ObsID $10206301nn$ and $10341701mm$, where $nn$ is either 01 or 02, and $mm$ ranges from 01 through 12. Together these data yield roughly $40$ ks of unfiltered exposure.

We processed the \nicer data using the $\textsc{nicerdas}$ version V004 within \textsc{heasoft} 6.24. Four epochs were defined (see Table \ref{tbl: timeline}) in which the source did not display rapid changes spectroscopically: epoch 1 is obs. 6 and 9; epoch 2 is obs. 26 and 27; epoch 3 is obs. 29-35; and epoch 4 is obs. 45-47. The data were filtered using standard cleaning criteria, i.e., a pointing offset $<54\arcsec$ from the \swift{}/XRT enhanced position, $>30\arcdeg$ from the dark Earth limb, $>40\arcdeg$ away from the bright Earth limb, and outside of the South Atlantic Anomaly. Additionally, we filtered out epochs of enhanced background, determined from the $12-15$~keV light curve \citep[see][for details]{Bult2018a,Bult2018b}. After filtering, we retained 26.7 ks of good time exposure. The $1-10$~keV background contribution to our observations was $0.6-1.5$ cts/s, as estimated from \nicer observations of blank field regions. For comparison, the source rate in this band varied between $\sim50-100$ cts/s.

\subsection{The Chandra X-ray Observatory (\chandra{})}
We observed \maxinos{} for 2.6~ks with the \cxo/High-Resolution Camera \citep[][HRC]{chandra_hrc_1} for best imaging resolution ($\sim{}\hspace{-1mm}0.4$\arcsec{}) and to avoid pileup. We used \texttt{ciao v4.9.3 repro} and \texttt{dmstat} commands to centroid the source with a 20 pixel radius. 

\subsection{The INTErnational Gamma-Ray Astrophysics Laboratory (\integral{})}
During its outburst, \maxinos{} was visible within the field of view of the \integral{} IBIS/ISGRI \citep{ubertini03,lebrun03} and the two JEM-X units \citep{lund03} from 2018 January 27 at 14:29 to 2018 April 11 at 11:00 (UT). Relevant publicly available data were collected during the satellite revolutions 1913-1919, 1922, 1926-1929, 1935, and 1937-1940. We analyzed all data by using version 10.2 of the \textsc{Off-line Scientific Analysis software} (OSA) distributed by the ISDC \citep{courvoisier03}. \integral{} observations are divided into ``science windows'' (SCWs), i.e., pointings with typical durations of $\sim$2 -- 3~ks. Only SCWs in which the source was located to within an off-axis angle of 4.0$^{\circ{}}$ from the center of the JEM-X field of view were included in the analysis. For IBIS/ISGRI, we retained all SCWs where the source was within an off-axis angle of 12$^{\circ{}}$ from the center of the instrument field of view.  

\subsection{The Neil Gehrels Swift Observatory/Ultra Violet Optical Telescope (\swift/UVOT)}
We utilized the \swift{}/UVOT \citep{swiftUVOT} UVW1, UVW2, and UVM2 filters, with central wavelengths of 2600, 1928, and 2246 \AA{}, respectively. Obs. 2, 3, and 4 were not utilized as \maxinos{} was off the chip.

For each observation, we created a circular source region (r=5\arcsec{}) centered at the best \cxo{} location ($\S{}$\ref{subsection: x-ray photometry}) and a circular background region (r=18\arcsec{}) near the source. We first retrieved and applied the aspect correction from the USNOB1 catalog using the \textsc{uvotskycorr ftool}. We then summed separate exposures with \textsc{uvotimsum} and used \textsc{uvotsource} to estimate source brightness at the 3$\sigma{}$ threshold. The source was not detected in any observations due to the heavy extinction in the Plane. The estimated $3\sigma$ upper limits are listed in the rightmost column of Table~\ref{tbl: Swift parameters}.

\begin{table}
	\begin{tabular}{lccc} 
		Obs. &Instrument/band & Magnitude & 3$\sigma{}$ lower lim.\\ 
        & & & Vega Mag. \\ \hline\hline
        10. & SIRIUS/J && $>$18.6  \\ 
        10. & SIRIUS/H && $>$18.0  \\ 
        10. & SIRIUS/K$_{S}$ && $>$17.1 \\ \hline\hline
	\end{tabular} 
    \caption{IR magnitudes and upper limits of J1621 counterparts. Obs. column is cross-referenced from Tbl. \ref{tbl: timeline}}  
    \label{tbl. IR+radio results}
\end{table}	

\subsection{The Gemini Observatory (\gemini{})}
We utilized the J, H, and K$_{s}$ filters on the FLAMINGOS-2 instrument \citep{gemini_flamingos2}, mounted on \gemini{} South. Using the \chandra{} localization, we were able to identify an IR counterpart to \maxinos{}. The source photometric variation and spectra are reported in Bahramian et al. (in preparation).

\subsection{The InfraRed Survey Facility (\irsf{})}
\irsf{} is a 1.4~m telescope in Sutherland observatory, South Africa. We used the simultaneous-imaging camera SIRIUS \citep{Nagashima1999,Nagayama2003} on \irsf{} \citep{Glass+Nagata2000}, which has a 7.7 square arcminute field of view. 
 We measured magnitudes of 2MASS \citep{Skrutskie2006} sources in the field of view for a calibrated search for a \maxinos{} counterpart in the J (1.25 $\mu{}$m), H (1.63 $\mu{}$m), and K$_{S}$ (2.14 $\mu{}$m) bands. Seeing during the 250~s (10~s x 25 frames) observations was limited to $\sim{}$2.5~arcsec, determined in the J band. The source was not detected in any of the observations (see Table \ref{tbl. IR+radio results}).

\subsection{The Australia Telescope Compact Array (\atca{})}
We observed \maxinos{} with \atca{} \citep{atca_upgrade} on 2017 November 18 between 19:00 UT and 24:00 UT, under project code CX399. 
We observed at central frequencies of 5.5 GHz and 9 GHz, each with 2 GHz of bandwidth in the 1.5C configuration. The flux and bandpass calibrator was PKS B1934$-$638. The complex gain calibrator was IERS B1600$-$489. The data were calibrated with the Multichannel Image Reconstruction, Image Analysis, and Display \citep[\textsc{MIRIAD,}][]{miriad} software package using the standard routines \citep{miriad}. We created Stokes I images using the \textsc{mfclean} procedure to properly account for the large fractional bandwidth at these relatively low central frequencies. \maxinos{} was not detected in either band. The flux density at the target location was $2.9\E{-5}$~Jy\,beam$^{-1}$ at 5.5 GHz and 2.2$\E{-5}$~Jy\,beam$^{-1}$ at 9 GHz.  The off-source rms was 2.4$\E{-5}$~Jy\,beam$^{-1}$ at 5.5 GHz and 1.4$\E{-5}$~Jy\,beam$^{-1}$ at 9 GHz.  To calculate the upper limit of the radio flux density, we took the measured Stokes I flux density at the target location and added it to 3$\times{}$ the off-source rms. 
The resulting upper limit values were $<0.10$~mJy at 5.5 GHz and $<0.064$~mJy at 9 GHz.  A bright, extended source approximately 27\arcmin{} to the east of \maxinos{} dominated the field at 5.5 GHz and contributed to the rms in the image, making it significantly higher than the theoretical thermal noise.


\section{Persistent X-ray emission} \label{analysis}
We discuss below the source localization, its persistent emission lightcurve and spectral evolution, and the results of our temporal analysis. Throughout our analyses, uncertainties are reported at the 90\% level unless otherwise specified. 

\subsection{X-ray Source Localization} \label{sec: localization} \label{sec: photometry}
\label{subsection: x-ray photometry}
We triggered our \cxo ToO and observed the \swift/XRT enhanced error box of the source \citep{swift_tile_ATEL,XRTenhanced1,XRTenhanced2} on 2018  February 22 for 2.6~ks with \cxo/HRC-I.  The source was seen with a net count rate of 1.77$\pm{}$0.03~cts s$^{-1}$ (S/N$>$55 using \texttt{celldetect}) at the aimpoint. We determined the position of \maxinos{} to be at RA, DEC (J2000) = 16$^{h}$ 20$^{m}$ 22$^{s}$.09, $-50^{\circ}\, 01\arcmin{}\, 09.39\arcsec{} \pm{}0.8\arcsec{}$; the total location uncertainty is dominated by the \chandra{} systematic pointing error. This is the best-known localization of the source to date.

\subsection{X-ray Light Curve}

Prior to the discovery of \maxinos, this field of the sky had only been observed four times with current X-ray instruments. A 4.6~ks archival \swift{}/XRT observation on 2007 February 6 (OBSID 00036140001) yielded a 3$\sigma{}$ upper limit of 1.3$\times{}10^{-12}$~erg~s$^{-1}$~cm$^{-2}$ in the 0.3--10~keV range. A \chandra{}/ACIS-S observation on 2008 May 28 for 1.6~ks (OBSID 09602) was also a non-detection with a 90\% confidence upper limit of 7.5$\times{}10^{-14}$~erg~s$^{-1}$~cm$^{-2}$. The most recent archival data were obtained with \swift{}/XRT (0.3--10~keV) on 2012 June 1 (OBSID 00042867001) and 2017 May 5 (OBSID 00087355001), which also yielded 3$\sigma{}$ upper limits of 8.4$\times{}10^{-12}$~erg~s$^{-1}$~cm$^{-2}$ and 1.3$\times{}10^{-12}$~erg~s$^{-1}$~cm$^{-2}$, respectively. The latter was obtained within the scope of our DGPS program. 


\begin{figure}[!b]
	\centering
    \includegraphics[width=0.99\linewidth]{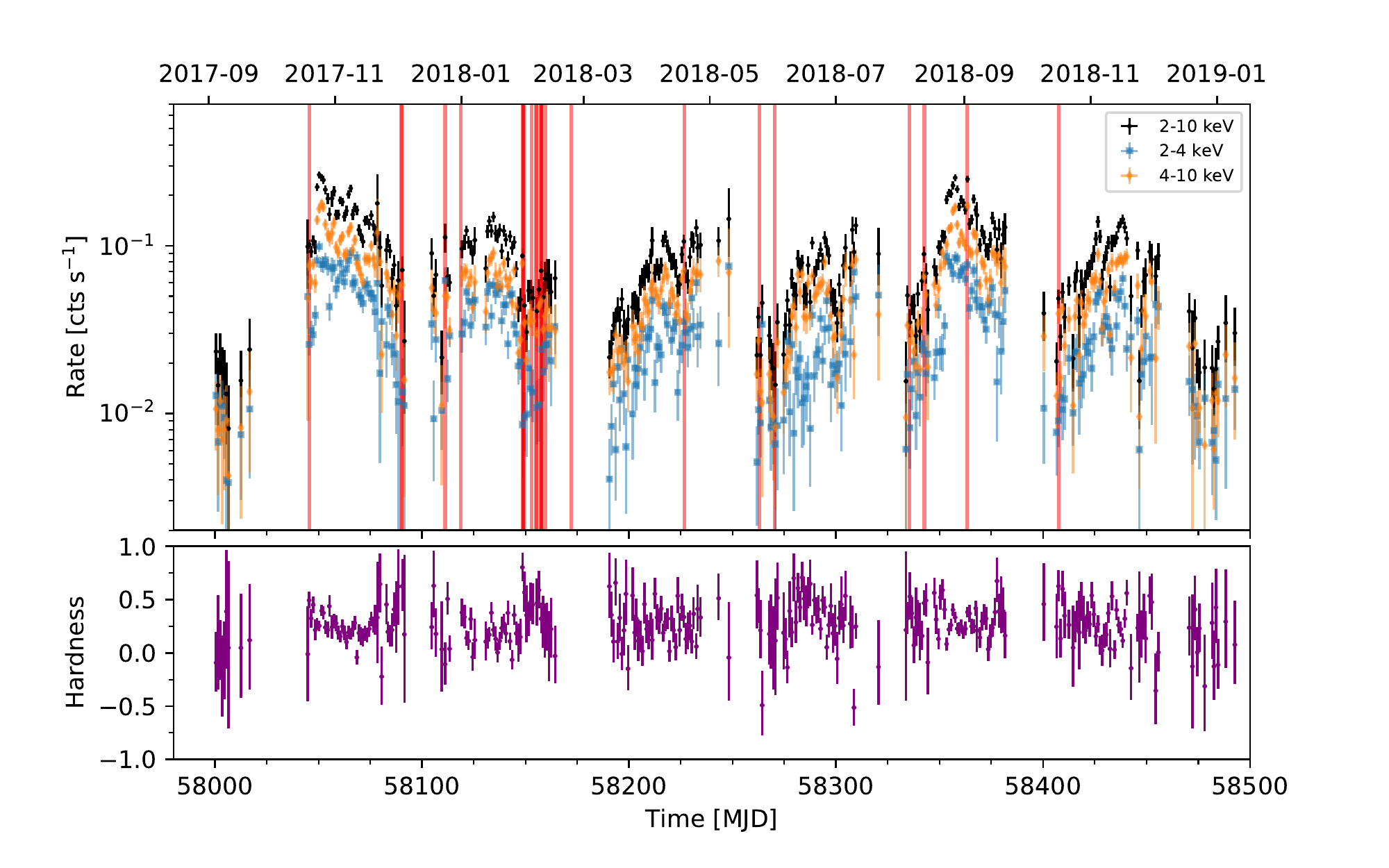}
	\caption{(\textit{Top}:) \maxi{} lightcurve of \maxinos{} plotted in the 2.0$-$10.0~keV (\textit{black points}), 2.0$-$4.0~keV (S, \textit{blue squares}), and 4.0$-$10.0~keV (H, \textit{orange diamonds}) bands are plotted. Red vertical lines indicate the observation time of an X-ray burst from any instrument. (\textit{Bottom}:) Hardness ratio = (H-S)/(H+S) for each point in the lightcurve. 
	}
    \label{fig: maxi lightcurve}
\end{figure}



\begin{figure}[!b]
	\centering
    \includegraphics[width=0.99\linewidth]{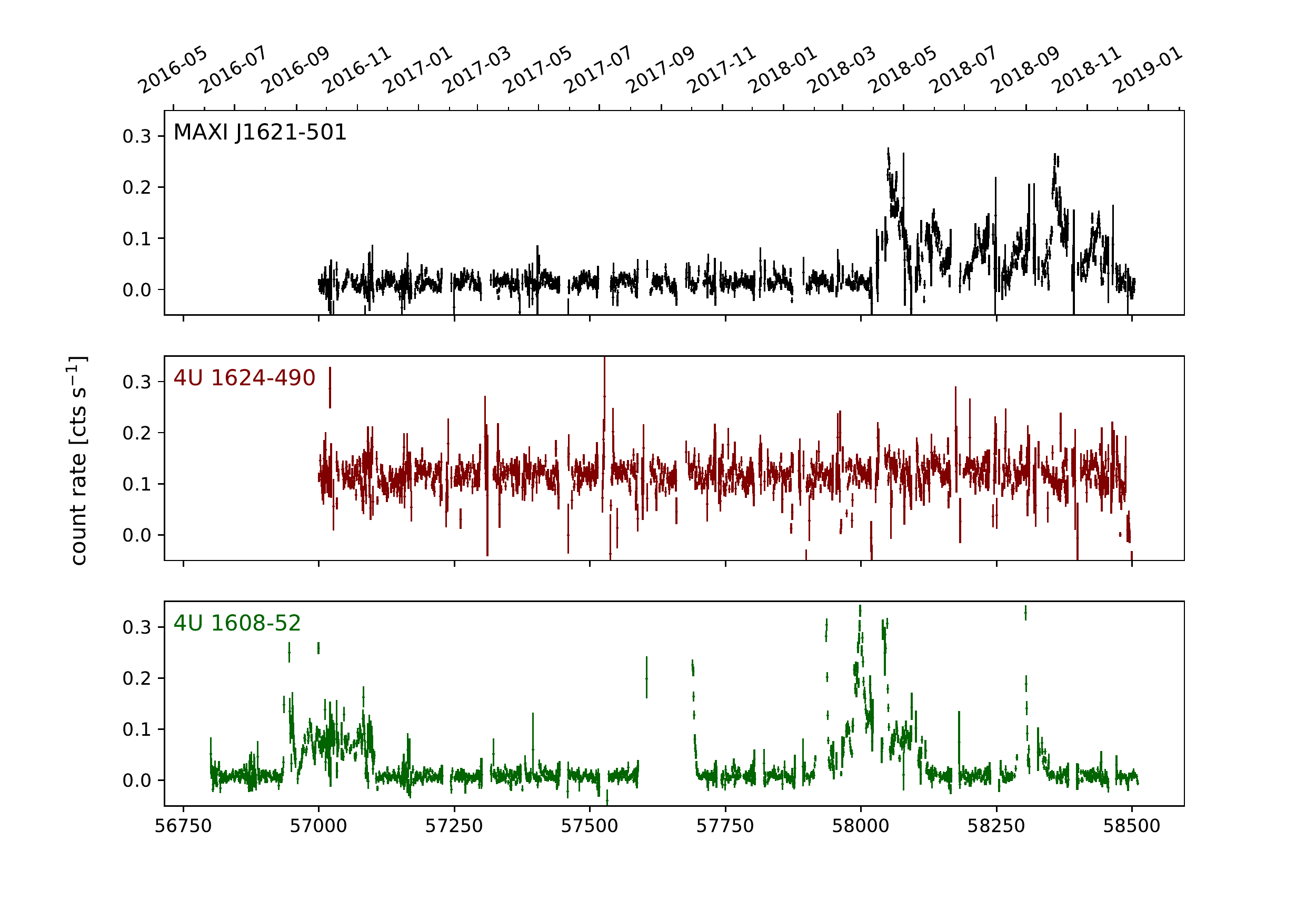}
	\caption{\maxi{} lightcurves of 3 sources in the vicinity of \maxinos{} do not display the $\sim78$ days modulation observed in the lightcurve of \maxinos{}. (\textit{Top to bottom}:) \maxinos{}, 4U 1624-490 (89\arcmin{} away), and 4U 1608-52 (161\arcmin{} away).}
    \label{fig: nearby sources}
\end{figure}


Following the source discovery, \maxi{} observed \maxinos{} continuously from 2017 October 19 until 2019 mid-February on a daily basis (excluding gaps due to Earth occultation and SAA passages). Fig. \ref{fig: maxi lightcurve} demonstrates the source flux variability, which appears to be episodic. Thus far we have identified six recurring episodes of activity, the primary peaks of which are separated by a $\sim78$ day interval, calculated by taking the arithmetic mean of the intervals between local maxima in Fig. \ref{fig: maxi lightcurve}. Although this is an intriguing feature, we noted its vicinity to the ISS precession period, so we took several steps to investigate its nature. The field on which \maxinos{} lies contains two other sources, which became active during the outburst of \maxinos{}. These sources are 4U 1624--490 a NS LMXB 90\arcmin{} away \citep{Christian+Swank1997} and 4U 1608--52 a NS LMXB 160\arcmin{} away \citep{Guver+2010}. We searched their light curves for similar modulations, assuming that if the \maxinos{} modulation is of instrumental nature and possibly associated with the ISS precession period, the same modulation will appear in their lightcurves. We plotted the image-fit data for all three sources (Fig. \ref{fig: nearby sources}); this method accounts for the contributions of the two nearby sources to the count rate of \maxinos. Although a significant modulation appears in the data of \maxinos{}, the other sources do not exhibit evidence for such an episodic activity. Barring unknown additional instrumental effects due to the ISS, we discuss in $\S{}$\ref{discussion} whether the \maxinos{} episodes are intrinsic source properties. 

One additional feature in the lightcurve is the appearance of the fourth episode at the same intensity as the first one; at the same time, the fifth episode exhibits a structure similar to that of the second. These similarities may indicate a possible secondary period of the order of 304 days; this claim, however, needs to be substantiated with longer observational intervals during a new source activation.

\begin{figure}[!ht]
	\centering
	\includegraphics[width=0.95\linewidth]{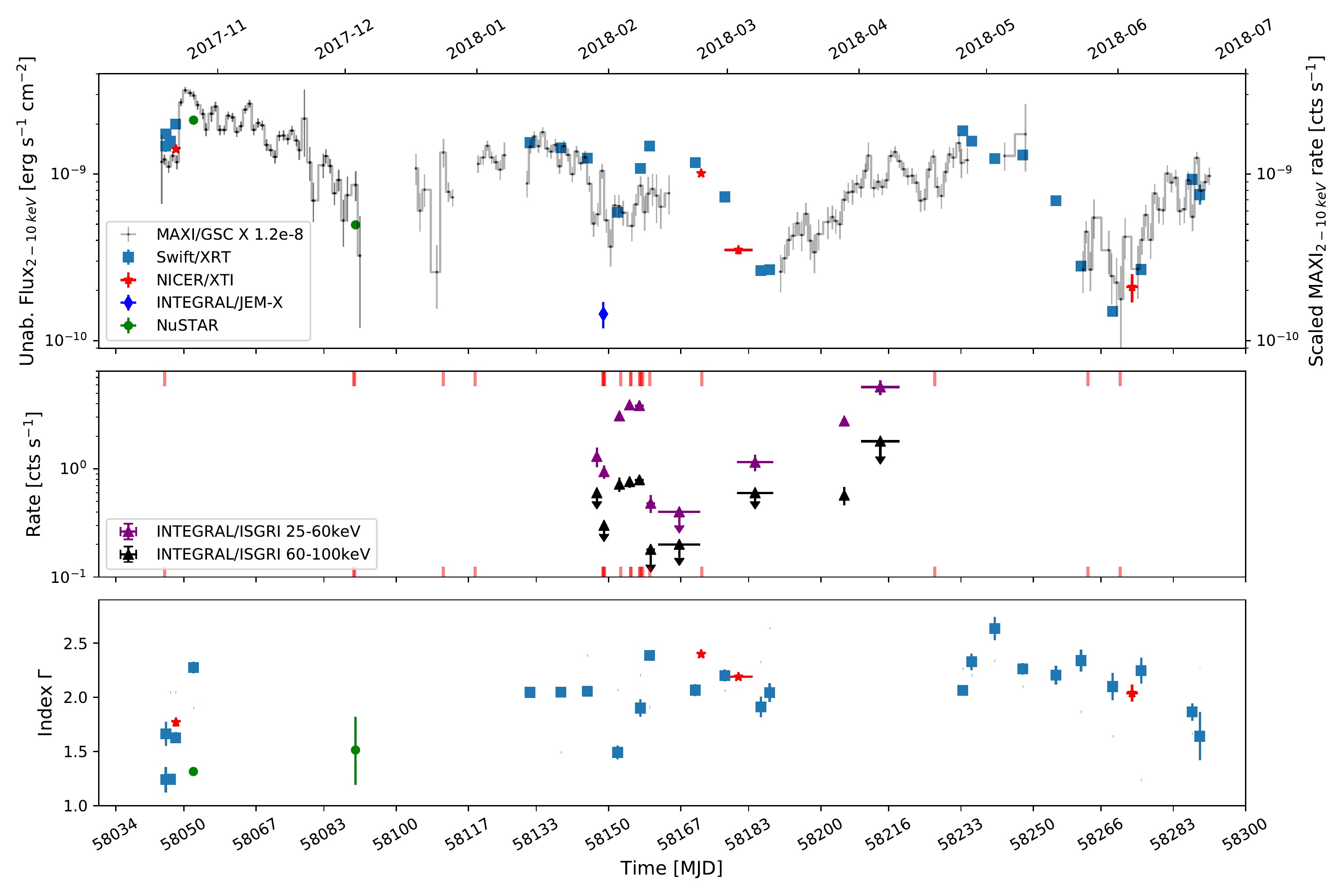}
	\caption {(\textit{Top:}) Multi-instrument unabsorbed flux evolution of \maxinos{} in the energy range of $2-10$\,keV. The \maxi{} count rate data are scaled by an arbitrary factor of 1.2\E{-8}. The \integral{}/JEM-X flux point was reported in the energy range 3-10~keV \citep{integral_detection}. (\textit{Middle:}) High-energy \integral{}/ISGRI data in two energy bands, red lines denote the 21 type I X-ray bursts within this interval (\textit{Bottom:}) Evolution of the powerlaw spectral index. Colors follow the legend in the top panel.}
    \label{fig. lightcurve}
\end{figure}

Fig. \ref{fig. lightcurve} combines the \maxi{} data with sporadic observations with \swift/XRT, \nicer/XTI, \nustar, and \integral/JEM-X and ISGRI. We note here that the MAXI data are in count rates and were scaled along the vertical axis to match the other instruments' observations, which are in flux units. Overall, there is good agreement across all instruments, corroborating that the superorbital modulations in the lightcurve are not instrumental in origin. We note that the two \integral/ISGRI lightcurves are also count rates in two energy bands ($25-60$~keV, $60-100$~keV) with their detection significance measured from the instrument mosaics extracted in different revolutions. The importance of this dataset is that it follows the \maxinos{} lightcurve rise in the X-ray (top panel) near MJD 58150. After this rise, the hard X-ray intensity drops rapidly near a peak of soft X-ray emission. The possible nature of this major dip is further discussed in \S{}\ref{discussion}. 



\subsection{X-ray Spectroscopy} \label{section:persistent x-ray spectroscopy}
Spectra of LMXBs are usually fit with a thermal component and a non-thermal component. In addition, the spectrum can contain a disk reflection component from ionized species in the accretion disk and Comptonization, produced by upscattering of the incident emission on a free electron halo. We discuss below our spectral fits to the data. Starting with \nustar{} and \nicer{}, we determined the continuum and narrow spectral features of \maxinos{}. The former mission has a high broadband sensitivity, and the latter has large effective collecting area and high spectral resolution.  These fits agree well with \swift/XRT observations, in spite of differences in spectral range and resolution. Finally, we show the contribution of \integral{} at energies up to $\sim{}$200~keV. To model contributions from the interstellar medium we used abundances from \citet{Wilms2000} and cross sections from \citet{Verner1996} in \texttt{xspec v12.10.0}. 


\begin{figure}[!h]
	\centering
    \includegraphics[width=0.49\linewidth]{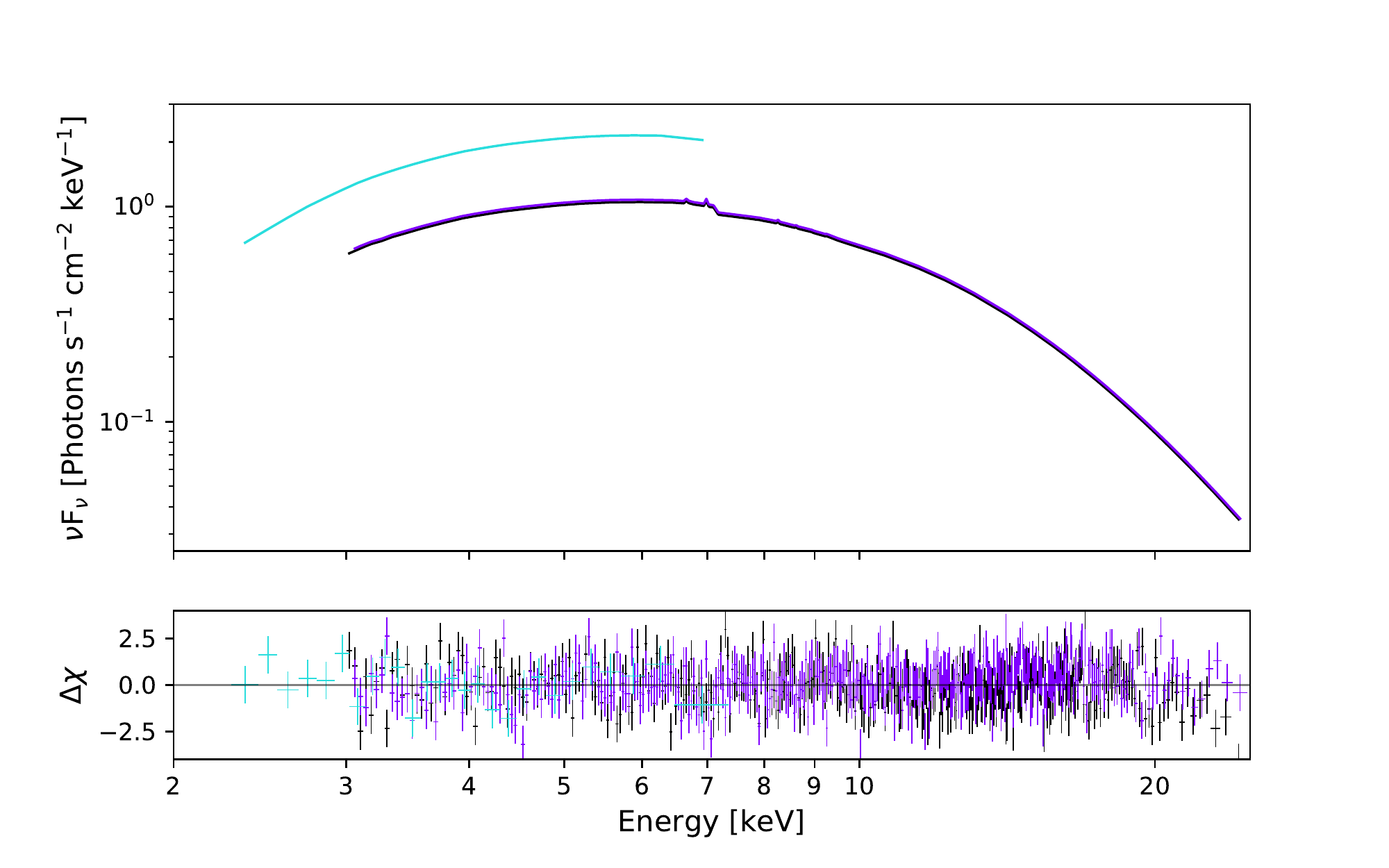}
	\includegraphics[width=0.49\linewidth]{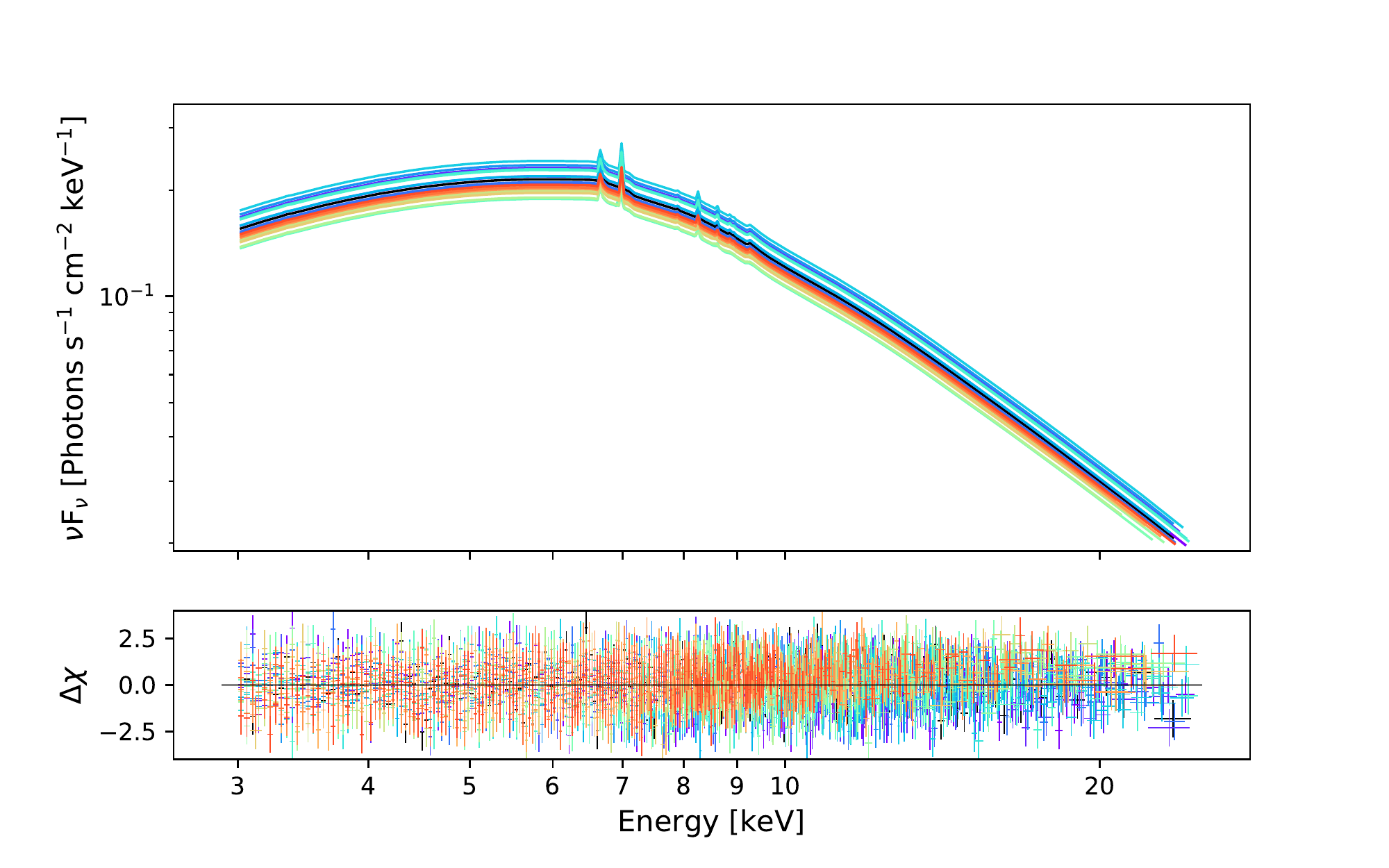}
	\caption{(\textit{Left:})\nustar{} and \swift{}/XRT (set 1) persistent emission spectra fit with an absorbed blackbody plus disk reflection model with a multiplicative constant. Only the model is shown. The bottom box shows the fit residuals in units of $\chi{}$. (\textit{Right:})  Similar for the remaining \nustar{} observations (set 2).}
	\label{fig: nustar spectrum background}
\end{figure}

\begin{table}[t]
\begin{tabular}{cccccc}
Epoch & nH                 & $\Gamma{}$ & Flux$_{2-10~keV}$ & $\chi{}^{2}$/dof & red. $\chi{}^{2}$ \\ 
      & 10$^{22}$cm$^{-2}$ &          & 10$^{-9}\;\frac{erg}{s\;cm^{2}}$  &              \\ \hline\hline

1 & 5.1 $\pm$ 0.1 & 1.77 $\pm$ 0.04 & 1.41 $\pm$ 0.02 & 744/655 & 1.14 \\
2 & 5.7 $\pm$ 0.1 & 2.40 $\pm$ 0.02 & 1.01 $\pm$ 0.01 & 911/676 & 1.35 \\
3 & 4.8 $\pm$ 0.3 & 2.19 $\pm$ 0.02 & 0.35 $\pm$ 0.02 & 958/717 & 1.34 \\
4 & 5.5 $\pm$ 0.2 & 2.04 $\pm$ 0.08 & 0.21 $\pm$ 0.04 & 687/599 & 1.15 \\ \hline\hline
\end{tabular}
\caption{The spectral parameters best fit to four epochs of \nicer{}/XTI data.}
\label{tbl: NICER spectral parameters}
\end{table}

\begin{table}
\begin{tabular}{ccc} 
		Set of spectra & set 1           & set 2 \\
			  & (obs. 12 \& 13) & (obs. 15 \& 16) \\ \hline\hline
		nH (E22 cm$^{-2}$)     & 4.23 $_{-0.29}^{+0.30}$ & 3.02 $_{-0.97}^{+0.89}$ \\  
		$\Gamma{}$      & 0.69 $_{-0.34}^{+0.42}$ & 1.75 $_{-0.34}^{+0.38}$ \\  
		Ecut [keV] & 2.77 $_{-0.64}^{+0.53}$  & 5.78 $_{-2.22}^{+2.62}$ \\  
		CPLnorm     & 0.75 $_{-0.11}^{+0.13}$ & 0.27 $_{-0.10}^{+0.12}$ \\  
		BB kT [keV]  & 2.32 $_{-0.09}^{+0.08}$ & 1.24 $_{-0.20}^{+0.13}$ \\
		BBnorm      & 1.70 $_{-0.83}^{+1.49}$ & 3.75 $_{-1.60}^{+2.05}$ \\  
		LineE [keV]  & 6.31 $_{-0.14}^{+0.14}$ & 6.19 $_{-0.18}^{+0.20}$ \\ 
		Width [keV]  & 3.15 $_{-0.45}^{+0.34}$ & 4.47 $_{-0.73}^{+0.40}$ \\ 
		Lnorm (E-3)     & 8.72 $_{-2.54}^{+2.47}$ & 7.76 $_{-3.03}^{+2.62}$ \\  \hline
		$\chi{}^{2}$ & 1051 with 877 bins (866 dof)  & 4865 with 4789 bins (4761 dof) \\ 
		\hline
		
		nH (E22 cm$^{-2}$)     & 4.39 $_{-0.14}^{+0.24}$ & 1.28  $_{-0.35}^{+0.49}$ \\  
		$\Gamma{}$      & 1.40 $_{-0.03}^{+0.05}$ & 1.51 $_{-0.32}^{+0.32}$ \\  
		Afe       & 0.50 $_{-0.50}^{+0.17}$ & 2.49 $_{-1.35}^{+2.56}$ \\  
		Ecut [keV] & 5.00 $_{-5.00}^{+0.04}$ & 7.94 $_{-0.69}^{+2.03}$ \\  
		logxi      & 4.05 $_{-0.02}^{+0.06}$ & 4.07 $_{-0.18}^{+0.30}$ \\  
		norm (E-3)      & 2.84 $_{-0.18}^{+0.20}$ & 0.44 $_{-0.12}^{+0.17}$\\  
		BB kT [keV]  & 1.79 $_{-0.03}^{+0.03}$ & 1.53 $_{-0.06}^{+0.03}$ \\ 
		norm      & 6.99 $_{-0.42}^{+0.39}$ & 3.61 $_{-0.43}^{+0.77}$ \\  \hline
		$\chi{}^{2}$ & 940 with 807 bins (797 dof) & 4852 with 4789 bins (4762 dof) \\ 
		\hline
		
        nH (E22 cm$^{-2}$)     & 3.65 $_{-0.39}^{+0.39}$ & 0.62 $_{-0.62}^{+0.65}$\\  
        T\_0 [keV] & 0.37 $_{-0.07}^{+0.05}$ &  0.36 $_{-0.26}^{+0.06}$\\  
        kT$_{e}$ [keV] & 2.51 $_{-0.05}^{+0.06}$ &  4.25 $_{-0.35}^{+0.51}$\\  
        Tau\_p &   13.12 $_{-0.79}^{+0.84}$ &  9.00 $_{-1.28}^{+1.50}$\\ 
        Compt norm (E-2)    & 36.44 $_{-6.13}^{+9.50}$ & 2.37 $_{-0.63}^{+1.07}$\\  
        BB kT [keV] & 1.33 $_{-0.04}^{+0.04}$ & 1.41 $_{-0.02}^{+0.02}$\\  
        BB norm     & 26.63 $_{-4.27}^{+4.90}$ & 7.49 $_{-0.69}^{+0.73}$ \\ \hline
        $\chi{}^{2}$ & 962 with 807 bins (798 dof) & 5002 with 4789 bins (4763 dof) \\ \hline\hline		

	\end{tabular} 
    \caption{Fits to \maxinos{} persistent emission, ({\it top}) absorbed cutoff power law plus blackbody plus lorentzian emission line ({\it middle}) absorbed xillver and blackbody ({\it bottom}) absorbed BB with Compton scattering from a halo of free electrons. Left column is set 1 and the right column is set 2, as described in the text.}  
    \label{tbl: nustar fit parameters}
\end{table}	

\subsubsection{\nustar{} \& \nicer{}}

Since the first \nustar{} observation (obs. 12 from Table \ref{tbl: timeline}) was contemporaneous with an XRT observation (obs. 13 from Table \ref{tbl: timeline}), they were fitted together to span a larger energy range and to better constrain $N_{\text{H}}$; we designate these two observations as set 1 (\nustar{} and \swift{}/XRT). The other two \nustar{} observations, which were taken one day apart, were first fitted separately, and the fit parameters were found to be consistent with each other. We, therefore, fitted them jointly to better constrain the spectral model parameters; these two observations (obs. 15 \& 16 from Table \ref{tbl: timeline}) we designate as set 2 (only \nustar{}). To fit set 1 and set 2, we used a calibration factor between each spectrum with the \nustar{} FPMA spectrum and the \nustar{} FPMA CHU 2 spectrum used as a reference (i.e., prefactor=1), respectively. The calibration factor was left free to vary in the other spectra. The value of the parameter had a maximum difference in set 1 of 12.7\% in the FPMA CHU12 spectrum (average of 6.7\%) and a maximum difference in set 2 of 104.1\%, in the XRT spectrum, reflecting the calibration difference between \nustar{} and \swift{}. 

Both sets of spectra (set 1, 2$-$25~keV, and set 2, 3$-$25~keV) were independently fitted with an absorbed blackbody (BB) plus a Comptonized emission component \citep{Titarchuk1994}. In set 1, the fit left a systematic residual pattern in the range of 6-7~keV, indicative of the presence of emission features. We then fitted this set with an absorbed BB plus a cutoff PL model, which resulted in even larger $\chi{}^{2}$ for set 1 and 2 (see Table \ref{tbl: nustar fit parameters}). This fit still left high systematic residuals at the same energy. Finally, we fitted a disk reflection model, \texttt{xillver} \citep{xillver_model}\footnote{Xillver is a subset of relxill: \href{http://www.sternwarte.uni-erlangen.de/~dauser/research/relxill/}{http://www.sternwarte.uni-erlangen.de/\~{}dauser/research/relxill/}}, paired with a BB (Fig. \ref{fig: nustar spectrum background}), which left no systematic residuals. The latter fit decreased the $\chi{}^{2}$ fit statistic by 22 ($\chi{}^{2}$=940)  and 150 ($\chi{}^{2}$=4852) from the Comptonized BB model for set 1 and set 2, respectively. Three parameters in the \texttt{xillver} model were frozen: redshift z=0.0, inclination angle Incl=30$^{\circ}$, and the reflected fraction refl\_frac=1.0 (100\% of intensity emitted towards the disk). The quality of data did not allow for good constraints on the inclination angle. Fig. \ref{fig: nustar spectrum background} shows the fits of both sets to the disk reflection plus BB model and their residuals; the fit parameters are shown in Table \ref{tbl: nustar fit parameters}.

We simulated the xillver and BB model (see \S{}\ref{section: spectral simulations}) and found that we could not reliably return the best fit parameters given the quality of our spectra. Along with setting the inclination angle, this informs us that this model cannot be adequately tested, so we restricted ourselves from then on to an absorbed cutoff PL plus BB plus a Lorentzian to model the contribution from a feature near 6.4~keV. This model resulted in reduced $\chi{}^{2}$=1.21 in set 1 and reduced  $\chi{}^{2}$=1.02 in set 2; it also allowed us to estimate the line continuum equivalent widths (0.27~keV and 1.43~keV, respectively) and fluxes (9.13$\times{}10^{-11}$~erg s$^{-1}$ cm$^{2}$ and 6.30$\times{}10^{-11}$~erg s$^{-1}$ cm$^{2}$, respectively). The fit parameters are recorded in Table \ref{tbl: nustar fit parameters}.

We split the \nicer data into four distinct epochs (plotted as red stars in Fig. \ref{fig. lightcurve}), where each epoch represents a closely-spaced set of observations (see $\S{}$\ref{sec:nicer data}). For each epoch we extracted the $1-10$~keV spectrum and fit it with several absorbed (\texttt{Tbabs}) models. We tried PL, disk blackbody (DBB), BB with Comptonization, and BB with a cutoff PL. Of these models, the PL consistently fit the best, with the DBB providing a worse fit (0.1 - 0.8 units of red $\chi{}^{2}$). The other models did comparably well but with more parameters and were thus discarded from further consideration.

The best fit PL parameters are shown in Table \ref{tbl: NICER spectral parameters}. We note here that of the four PL spectral indices ($\Gamma{}=$1.77$\pm{}$0.04, 2.40$\pm{}$0.02, 2.19$\pm{}$0.02, and 2.04$\pm{}$0.08), three are above 2 and one is 1.77. The softer spectra seem to all have occurred during episodic minima, while the hardest of the four was measured during the ascending part of the first episode. The reduced $\chi{}^{2}$ values (1.1, 1.3, 1.3, 1.2) mostly reflect systematic residuals around 1.7~keV and 2.1~keV, both of which are known instrumental features. 

\begin{figure}[!b]
\centering
\includegraphics[scale=1]{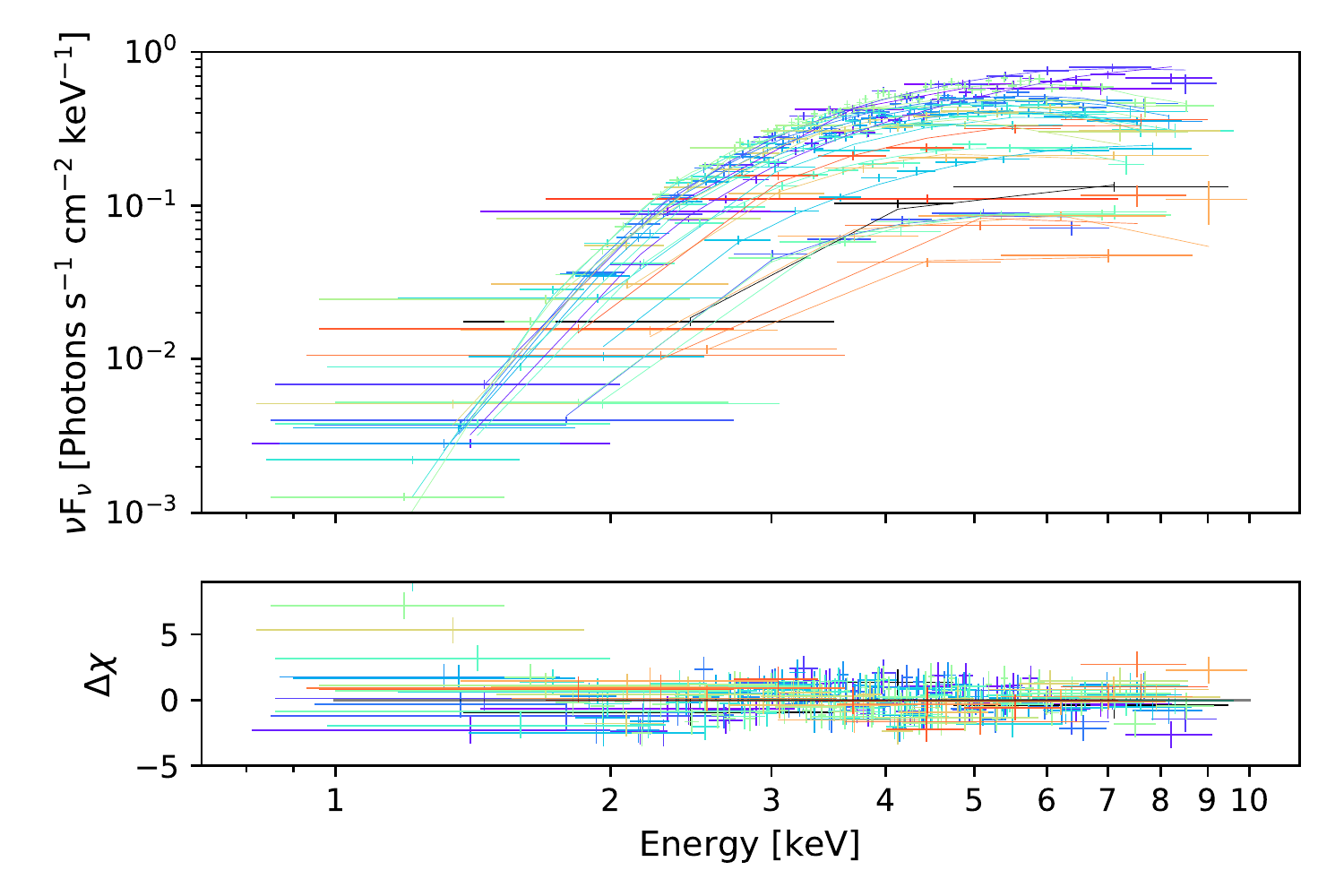}
\caption{\swift/XRT data jointly fitted with an absorbed PL model. Data are binned to a minimum of 15 photons for ease of viewing.}
\label{fig: all swift PL}
\end{figure}

\begin{table}
\begin{center}
\begin{tabular}{lcccc} \\
\swift{} OBSID &  Power law $\Gamma{}$   & Disk BB kT & UVOT filter & UVOT 3$\sigma{}$ upper limit\\ 
               &              & keV&             & 10$^{-17}$erg~s$^{-1}$~cm$^{-2}$~\AA{}$^{-1}$\\ \hline\hline
00036140001 &  -      &  -      & UVW1 & 3.49 \\
00042867001 &  -      &  -      & UVM2 & 4.84 \\
00087355001 &  -      &  -      & UVW1 & 1.13 \\
00010352001 & 1.24$_{-0.20}^{+0.20}$ & 5.99$_{-2.40}^{+46.88}$ &  -&  -\\
00010357001 & 1.67$_{-0.18}^{+0.18}$ & 2.86$_{-0.46}^{+0.73}$ &  -&  -\\
00036140002 & 1.24$_{-0.06}^{+0.06}$ & 5.29$_{-0.64}^{+0.93}$ & UVM2 & 4.62 \\
00036140003 & 1.63$_{-0.08}^{+0.08}$ & 3.08$_{-0.25}^{+0.31}$ &  -&  -\\
00036140004 &  -      &  -      & UVM2 & 4.23 \\
00087355002 & 2.28$_{-0.09}^{+0.09}$ & 1.85$_{-0.10}^{+0.11}$ & UVW1 & 5.83 \\
00036140006 & 2.05$_{-0.07}^{+0.07}$ & 2.09$_{-0.09}^{+0.10}$ & UVW1 & 4.54 \\
00036140007 & 2.05$_{-0.07}^{+0.07}$ & 2.13$_{-0.09}^{+0.10}$ & UVM2 & 4.42 \\
00036140008 & 2.06$_{-0.07}^{+0.07}$ & 2.09$_{-0.09}^{+0.10}$ & UVW2 & 4.57 \\
00036140009 & 1.49$_{-0.11}^{+0.11}$ & 3.57$_{-0.42}^{+0.59}$ &  -&  -\\
00036140011 & 1.90$_{-0.14}^{+0.14}$ & 2.47$_{-0.25}^{+0.33}$ & UVW2 & 4.55 \\
00036140012 & 2.39$_{-0.07}^{+0.07}$ & 1.69$_{-0.06}^{+0.07}$ & UVW1 & 4.57 \\
00036140014 & 2.07$_{-0.09}^{+0.09}$ & 2.10$_{-0.12}^{+0.14}$ & UVM2 & 5.49 \\
00036140016 & 2.20$_{-0.09}^{+0.09}$ & 1.92$_{-0.11}^{+0.12}$ & UVW2 & 4.33 \\
00036140017 & 1.91$_{-0.16}^{+0.16}$ & 2.35$_{-0.27}^{+0.35}$ & UVW2 & 4.82 \\
00036140018 & 2.04$_{-0.14}^{+0.14}$ & 2.12$_{-0.19}^{+0.24}$ & UVW1 & 3.63 \\
00010670001 & 2.06$_{-0.05}^{+0.05}$ & 2.12$_{-0.06}^{+0.07}$ & UVW2 & 7.31 \\
00010670002 & 2.33$_{-0.13}^{+0.13}$ & 1.82$_{-0.13}^{+0.16}$ & UVW1 & 3.36 \\
00010670003 & 2.64$_{-0.18}^{+0.18}$ & 1.50$_{-0.12}^{+0.15}$ & UVW2 & 4.55 \\
00010670004 & 2.26$_{-0.09}^{+0.09}$ & 1.84$_{-0.09}^{+0.10}$ & UVW1 & 5.16 \\
00010670005 & 2.21$_{-0.14}^{+0.14}$ & 1.95$_{-0.16}^{+0.20}$ & UVW1 & 3.88 \\
00010670006 & 2.34$_{-0.18}^{+0.18}$ & 1.89$_{-0.19}^{+0.24}$ & UVW2 & 4.69 \\
00010670007 & 2.10$_{-0.20}^{+0.20}$ & 2.06$_{-0.26}^{+0.35}$ & UVW2 & 4.57 \\
00010670008 & 2.25$_{-0.21}^{+0.21}$ & 2.04$_{-0.26}^{+0.36}$ & UVW1 & 3.28 \\
00010670010 & 1.87$_{-0.14}^{+0.14}$ & 2.62$_{-0.28}^{+0.37}$ & UVW1 & 3.81 \\
00010670011 & 1.64$_{-0.37}^{+0.37}$ & 3.04$_{-0.88}^{+2.80}$ & UVW2 & 9.35 \\ \hline{} 
N$_{\text{H}}\:[10^{22}$~cm$^{-2}$] & $5.53^{+0.10}_{-0.10}$ & 4.22$^{+0.07}_{-0.06}$ & &\\
\hspace{0.5cm} red $\chi{}^{2}$ & 3170/2264=1.40 & 3056/2264=1.35 & & \\ \hline\hline
\caption{The parameters for the best fits to the absorbed PL and absorbed disk BB models. N$_{\text{H}}$ was linked between all observations. The UVOT 3$\sigma{}$ upper limits are reported in the rightmost column, uncorrected for extinction.}
\renewcommand\thetable{2}
\label{tbl: Swift parameters}
\end{tabular}
\end{center}
\end{table}

\subsubsection{\swift{}/XRT}

We loaded all XRT spectra (extracted and grouped) using \texttt{pyxspec} and fitted them jointly with multiple functions, including BB, disk reflection, and non-thermal models. Of these, the best fits were provided by an absorbed disk-BB (DBB) and a single absorbed PL model (Fig. \ref{fig: all swift PL}), with $\chi{}^{2}$ = 3056 and 3170 for 2264 dof, respectively. Unlike with the \nicer{} fits, both models fit the \swift/XRT data equally well, and we report these results in Table \ref{tbl: Swift parameters}.  We first fitted all observations keeping all parameters free to vary. The resulting $N_{\text{H}}$ values were in the 90\% confidence interval range (4.09$_{-0.86}^{+0.97}$ -- 6.13$^{+0.35}_{-0.34}$)$\times10^{22}$ cm$^{-2}$. We then linked $N_{\text{H}}$ between all observations; these fits resulted in $N_{\rm H}$ of $5.53^{+0.10}_{-0.10}\times10^{22}$ cm$^{-2}$ and $ 4.22^{+0.07}_{-0.06}\times10^{22}$ cm$^{-2}$, for the PL and DBB, respectively.

The DBB fits the \swift/XRT data slightly better, however, we utilized the PL spectral parameters for comparison with those of the \nustar{} data. We found that the \swift/XRT best-fit spectral indices, $\Gamma{}$, show random variability between 1.24 $\pm{}$ 0.20 and 2.64 $\pm{}$ 0.18 (Fig. \ref{fig. lightcurve}, bottom panel). Three data sets show high, positive residuals below 2~keV. This could be due to the presence of a low-energy, narrow band component, which we attempted to fit with a second BB; other LMXB spectra have been successfully fit with multi-temperature blackbody models, usually attributed to the accretion disk (the ``Eastern Model" of \citet{Mitsuda+1989}). The result was a lower fit statistic but unconstrained fit parameters, since the component lies near the edge of the spectral band. This BB component was consequently dropped from the model. 
\begin{figure}[!b]
	\centering
	\includegraphics[width=0.6\linewidth]{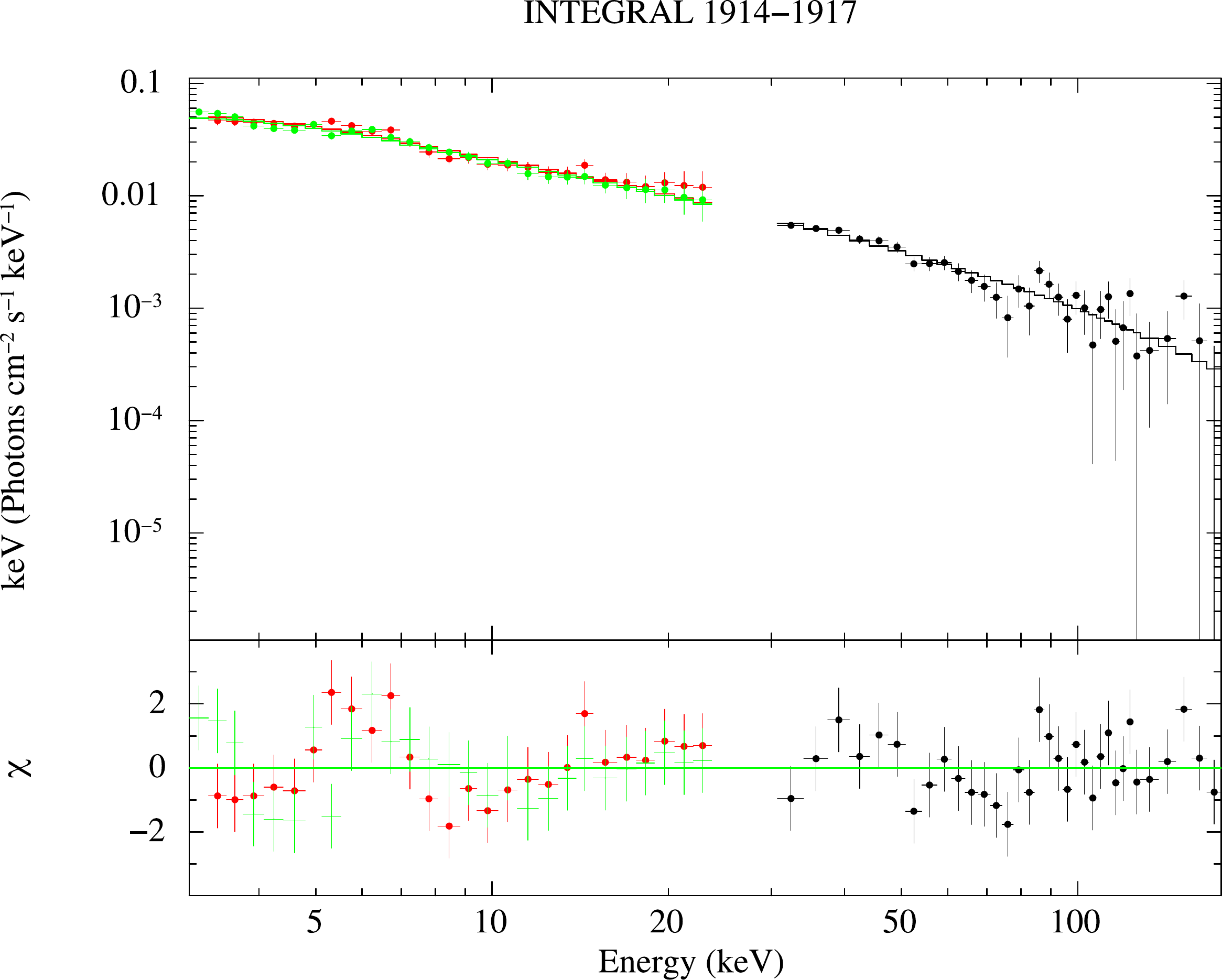}
	\caption{The \integral{} spectra extracted from the combined data in revolutions 1914-1917. The ISGRI data are in black, the JEM-X1 data in red, and the JEM-X2 data in green. The best fit model is obtained with a cut-off power-law and the residuals from the best fit are reported in the bottom panel of the figure.}
\label{fig:integralspe}
\end{figure}

\begin{table}[!ht]
\centering
{\normalsize \begin{tabular}{lccccccc}
Revolution & Time span & Cts/s & Det. sign. & Cts/s & Det. sign. & Hardness ratio \\ 
& (MJD) & (25--60 keV) & (25--60 keV) & (60--100 keV) & (60--100 keV) & ($\sigma$) & \\ \hline\hline
1913 & 58146.81--58147.72 & 1.30$\pm$0.27 & 4.8 & $<$0.6 & --- & --- \\
1914 & 58148.26--58149.65 & 0.94$\pm$0.13 & 7.0 & $<$0.3 & --- & --- \\ 
1915 & 58152.13--58153.04 & 3.08$\pm$0.16 & 18.9 & 0.72$\pm$0.11 & 6.4 & 0.23$\pm$0.04\\
1916 & 58154.40--58155.53 & 3.89$\pm$0.14 & 28.8 & 0.76$\pm$0.09 & 8.2 & 0.20$\pm$0.02 \\
1917 & 58156.24--58158.34 & 3.84$\pm$0.09 & 40.9 & 0.79$\pm$0.06 & 12.3 & 0.21$\pm$0.02 \\
1918 & 58158.90--58160.99 & 0.48$\pm$0.09 & 5.7 & $<$0.18 & --- & --- \\
1919-1922 & 58161.72--58171.63 & $<$0.40 & --- & $<$0.20 & --- & --- \\
1926-1929 & 58180.26--58188.78 & 1.15$\pm$0.20 & 5.7 & $<$0.6 & --- \\
1935 & 58204.87--58206.18 & 2.76$\pm$0.17 & 16.2 & 0.57$\pm$0.11 & 5.1 & 0.21$\pm$0.11 \\
1937-1940 & 58209.45--58218.59 & 5.70$\pm$0.88 & 6.5 & $<$1.8 & --- & --- \\ \hline\hline
\end{tabular} } 
\caption{Count rates and hardness ratios obtained from all IBIS/ISGRI data collected during the outburst of J1621. Values of the count-rate preceeded by $<$ correspond to 3$\sigma$ upper limits. For comparison, the count-rates of the Crab in the 20$-$60~keV and 60$-$100~keV energy bands are 99.4$\pm$0.2~cts~s$^{-1}$ and 25.6$\pm$0.1~cts~s$^{-1}$, respectively (we used the publicly available observations of the Crab carried out during the satellite revolution 1921 for a total of 45~ks)}
\label{tbl: integral1}
\end{table}

\vspace{2cm}
\subsubsection{\integral{} \& \maxi{}}

We extracted the \integral{} IBIS/ISGRI mosaics in the $25-60$~keV and $60-100$~keV energy bands and inspected the detection significance of the source in these mosaics and the correspondingly measured count rates by IBIS/ISGRI in order to search for possible spectral variations. We show the results of this analysis in Table~\ref{tbl: integral1}. As no significant variations in the source hardness ratio were measured, we extracted two spectra: the first summed up all data in revolutions 1914--1917 to obtain the best signal-to-noise ratio, and the second used the data in revolution 1935, which is separated by the other revolutions by slightly more than 20~days but characterized by a relatively high source detection significance. We followed the same strategy for the extraction of the JEM-X1 and JEM-X2 spectra. In all cases, we removed from the data 1~ks of exposure around the 11 type I bursts detected with JEM-X in order not to contaminate the spectrum of the persistent emission.  

The combined ISGRI+JEM-X spectra from revolutions 1914-1917 (effective exposure time of 60.8 ks for ISGRI and 4.7 ks for each of the two JEM-X) could be well described ($\chi^2_{\rm red}$=1.06=87/82) by using a cut-off PL model. We fixed for all fits to the \integral{} spectral data the value of the absorption column density at 2.5$\times$10$^{22}$~cm$^{-2}$, as they were not sensitive to variations of this parameters within a factor of few from this value. We measured in this case a PL photon index of 1.96$\pm$0.06, a cut-off energy of 100$^{+40}_{-24}$~keV, and a $3-100$ keV flux of 1.03$\times$10$^{-9}$~erg~cm$^{-2}$~s$^{-1}$. The normalization constants introduced to take into account the inter-calibrations between the different \integral{} instruments were all compatible with unity (within the associated uncertainties). The statistics of the data collected in revolution 1935 (effective exposure time of 60.8 ks for ISGRI and 4.7 ks for each of the two JEM-X) are significantly lower than that in revolution 1914-1917 and thus a simple absorbed PL model can describe these data well ($\chi^2_{\rm red}$=1.00=16/16). We measured in this case a photon index of 1.3$\pm$0.5. If a cut-off PL model is used for the fit and the cut-off energy is fixed to the above value of 100~keV (resulting in $\chi^2_{\rm red}$=1.00=16/16), then the photon index of the PL in revolution 1935 would be 1.0$\pm$0.5, i.e., slightly harder than that measured for the revolutions 1914-1917 (we fixed in all cases the absorption column density at 2.5$\times$10$^{22}$~cm$^{-2}$). We show, as an example, the ISGRI+JEM-X spectra of the source obtained from the revolutions 1914-1917 data in Fig.~\ref{fig:integralspe}, together with the best fit model and the residuals from the fit. Although of low significance, we note the presence of large residuals at $\sim6$\,keV, in accordance with the results in the \nustar{} data. 


Finally, we used the \maxi{} data to obtain the source count rate hardness evolution, defined as the hard band minus the soft band divided by the sum of the bands, during MJD $58000-58492$ (Fig. \ref{fig: maxi lightcurve} (\textit{bottom})). We excluded time bins with count rates below 0~cts~s$^{-1}$ (due to background subtraction) and time bins where the absolute value of the hardness ratio plus error is larger than 1. Fig. \ref{fig. lightcurve} (\textit{bottom}) shows the PL indices of the spectral fits of the \swift/XRT, \nustar, and \nicer data points.  

\begin{table}[!t]
\centering
{\normalsize \begin{tabular}{lcccccc}
Model	&	Parameter	&	Input	&	Min Value	&   Max Value	&	Conf. Interval [\%]	&	Includes input	\\ \hline\hline{}
CPL+BB+L	&	nH (E+22 cm$^{-2})$	&	4.23	&	3.84	&   4.58	&	69	&	\checkmark \\  
	&	$\Gamma{}$	&	0.69	&	0.57	&	0.79	&	69	&	\checkmark\\  
	&	Ecut [keV]	&	2.77	&	2.55	&	2.96	&	70	&	\checkmark\\  
	&	CPLnorm	&	0.75	&	0.61	&	0.89	&	70	&	\checkmark\\  
	&	kT [keV]	&	2.32	&	2.23	&	2.37	&	72	&	\checkmark\\
	&	BBnorm	&	1.70	&	1.21	&	2.91	&	69	&	\checkmark\\  
	&	LineE [keV]	&	6.31	&	6.18	&	6.45	&	68	&	\checkmark\\ 
	&	Width [keV]	&	3.15	&	2.43	&	3.58	&	69	&	\checkmark\\ 
	&	Lnorm (E-3)	&	8.72	&	5.39	&	12.09	&	69	&	\checkmark\\  \hline

Xillver+BB	&	nH (E22 cm$^{-2}$)	&	4.39	&	4.28	&	4.61	&  70	&	\checkmark\\ 
	&	$\Gamma{}$	&	1.40	&	1.39	&	1.46	&	71	&	\checkmark\\ 
	&	Afe 	&	0.50	&	0.58	&	0.94	&	72	&	$\times{}$\\ 
	&	Ecut [keV]	&	5.00	&	5.04	&	5.23	&	69	&	$\times{}$\\ 
	&	logxi	&	4.05	&	4.05	&	4.15	&	70	&	\checkmark\\ 
	&	Xnorm (E-3)	&	2.84	&	2.75	&	3.01	&	71	&	\checkmark\\ 
	&	kT [keV]	&	1.79	&	1.77	&	1.81	&	69	&	\checkmark\\ 
	&	BBnorm	&	6.99	&	6.91	&	7.67	&	70	&	\checkmark\\  \hline \hline
\end{tabular} } 
\caption{Simulation results (\textit{top}) using the absorbed Xillver + BB model (\textit{bottom}) using the absorbed cutoffPL plus BB plus Lorentzian model. Column 7 denotes if the confidence interval from column 6 includes the input value used to create the synthetic spectra.}
\label{tbl: sim_results}
\end{table}

\subsubsection{Spectral Simulations} \label{section: spectral simulations}

We performed extensive simulations following the procedure described in Appendix A.1. of \citet{Guiriec+2013} for the two best spectral models, xillver+BB and cutoff PL+BB+Lorentzian. By performing these simulations, we tested our ability to recover the accurate spectral parameters, i.e., those used to create the simulations. For each model we produced 10$^{5}$ synthetic spectra (using model parameters from Table \ref{tbl: nustar fit parameters}) with the \textsc{fakeit} command in \textsc{xspec v12.10.0}; each synthetic spectrum was fitted with the same model used to produce it. For each parameter, we expected the probability distribution function (pdf) to peak close to the parameter value used to produce the synthetic spectrum. To reduce computational time, we synthesized spectra using the \nustar{} aspect correction, the background spectrum, rmf, and arf files from CHU2 from the \nustar{} FPMA in observation 15. The other spectra vary only by a multiplicative constant, which reflects the different combinations of CHU and FPMs.

The pdf of each parameter for the models discussed above are plotted in Figs. \ref{fig: dim_xillver_corner} and \ref{fig: bright_pl+bb+lorentz_corner}. However, some parameters of the xillver+BB model showed evidence of jumps (discontinuities). In the Fe abundance ($Af_{e}$) and E$_{\rm cut}$, this is indicative of the grid of models not encompassing enough of the parameter space, shown by excess at the edge(s) of the distributions, which are otherwise relatively smooth (Fig. \ref{fig: dim_xillver_corner}). In probability distributions with this issue, we truncated the distribution by removing these bins. We then redistributed the probability in the removed bins to the remaining smooth distribution, weighted with respect to the remaining bins' probability.

For both models, we calculated for each pdf the minimum and the maximum parameter values, which enclose the $\sim{}$68\% confidence interval as follows. From either side of each parameter probability distribution (rightmost tiles in Fig. \ref{fig: dim_xillver_corner} and \ref{fig: bright_pl+bb+lorentz_corner}), we calculated the cumulative distribution until its value surpassed 0.16. We then chose the previous half bin that did not surpass this value in order to denote the beginning of the $\sim{}$68\% confidence interval; the resulting confidence intervals are reported in column 6 of Tbl. \ref{tbl: sim_results}. These results tend to favor the Cutoff PL plus BB plus Lorentzian model because: a) all input parameters were recovered within the central $\sim{}$68\% confidence interval, and b) the parameter probability distributions were smooth, indicating an adequately broad grid of parameter values.



\subsection{X-ray Timing}

\begin{figure}[!b]
	\centering
	\includegraphics[scale=0.7]{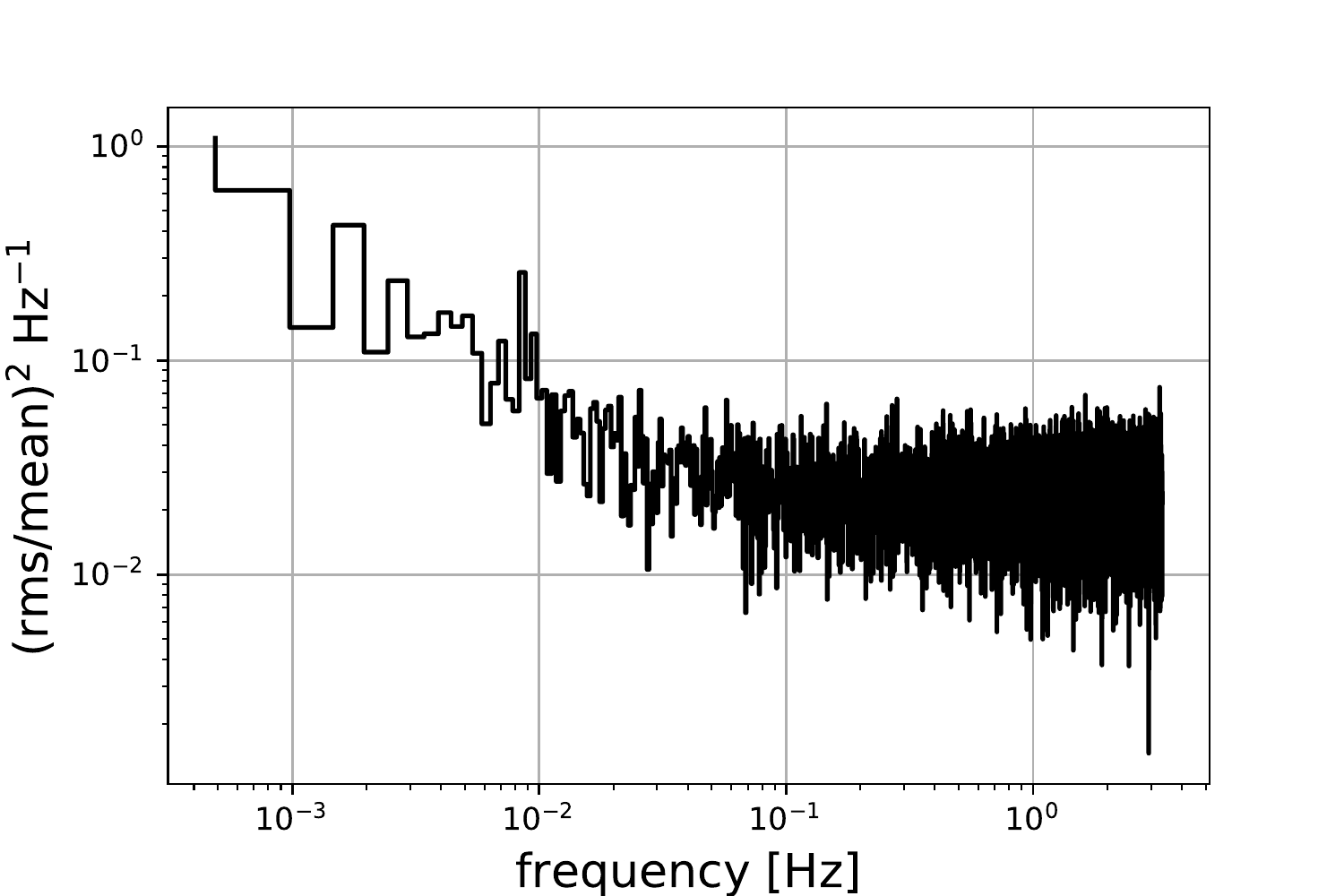} 
	\caption{Averaged periodogram of the \nustar{} data from Oct. 26. There is a candidate QPO at $\nu = 0.0088\,\mathrm{Hz}$, but its significance is low ($p=0.05$ corrected for $6826$ frequency trials).}
	\label{fig: NuSTAR powspec}
\end{figure}

\begin{table}[!t]
\centering
\caption{Parameter definitions and prior probability distributions for the power-law+constant model used to fit the periodogram and simulate data sets from the null hypothesis.}
{\normalsize \begin{tabular}{lccc}
Parameter & Definition & Prior Probability Distribution \\ \hline \hline
$\Gamma$ & power law index &  $\mathrm{Uniform}(0,5)$ \\
$\log{A_\mathrm{PL}}$ & power law amplitude & $\mathrm{Uniform}(-20,20)$ \\
$\log{A_{\mathrm{noise}}}$ & Poisson noise amplitude &  $\mathrm{Uniform}(-10,10)$ \\
\end{tabular} } 
\label{tbl:periodogram_priors}
\end{table}

We searched the second \nustar{} observation (Obs. 15, Table \ref{tbl: timeline}) for quasi-periodic oscillations in the source using the \textsc{stingray} timing package \citep{stingray, Huppenkothen+2019}, on a lightcurve spanning 19~ks ($3-79$\,keV). The data were first barycentered in the \textsc{nuproducts} routine using the operations-provided orbit file and the source centroided coordinates.

We produced an averaged periodogram using segments of $2048\,\mathrm{s}$ duration. The final periodogram includes 7 individual segments averaged together, utilizing all contiguous GTI intervals longer than the segment length. Because of the source's brightness, the \nustar{} data were strongly affected by dead time \citep{bachetti2015}. We corrected the periodograms of individual segments using the Fourier Amplitude Differencing (FAD) technique of \citet{bachetti2017}. In short, the FAD method utilizes the light curves of the two different detectors onboard \nustar{} to compute the difference of the Fourier amplitudes in the two detectors, which can be used to separate intrinsic source variability from the frequency-dependent effects of dead time. After correction, the segments were averaged together to produce a dead time-corrected averaged periodogram (Fig. \ref{fig: NuSTAR powspec}). 

We used the method laid out in \citet{vaughan2010} to search for narrow quasi-periodic signals in the averaged periodogram. We modeled the periodogram with a PL plus a constant to account for the white noise level, using fairly wide, uninformative priors for the parameters (Table \ref{tbl:periodogram_priors}). We first fitted the model to the data and computed the maximum outlier in the residuals. Subsequently, we used Markov Chain Monte Carlo (MCMC) implemented in the Python package \texttt{emcee} \citep{foremanmackey2013} to sample the parameter space of the models. We then generated simulated periodograms from random samples from the posterior probability distribution. For each, we fitted a power law model, and computed the highest outlier in the residuals of these simulated periodograms. We then compared the highest outliers derived from the simulated periodograms according to our null hypothesis (no signal) to the highest outlier in the observed periodogram. 

There is a potential candidate detection of a narrow quasi-periodic signal at $\nu = 0.0088\,\mathrm{Hz}$, or a period of $P = 113\,\mathrm{s}$. However, the trial-corrected significance is only $p=0.05$, indicating that this signal could potentially be explained by noise. In order to independently confirm the signal, we also searched \swift{} obs. 13 for a signal at the same frequency and found no trace of a similar QPO in this data set. However, it is important to note that the \swift{} dataset was much shorter ($566\,\mathrm{s}$ total duration) and was heavily affected by pile-up, with only a fraction of the photons actually recorded. It is, therefore, possible that the lack of signal in the \swift{} data could be related to the data quality.

\begin{figure}[!b]
	\centering
	\includegraphics[width=0.32\linewidth]{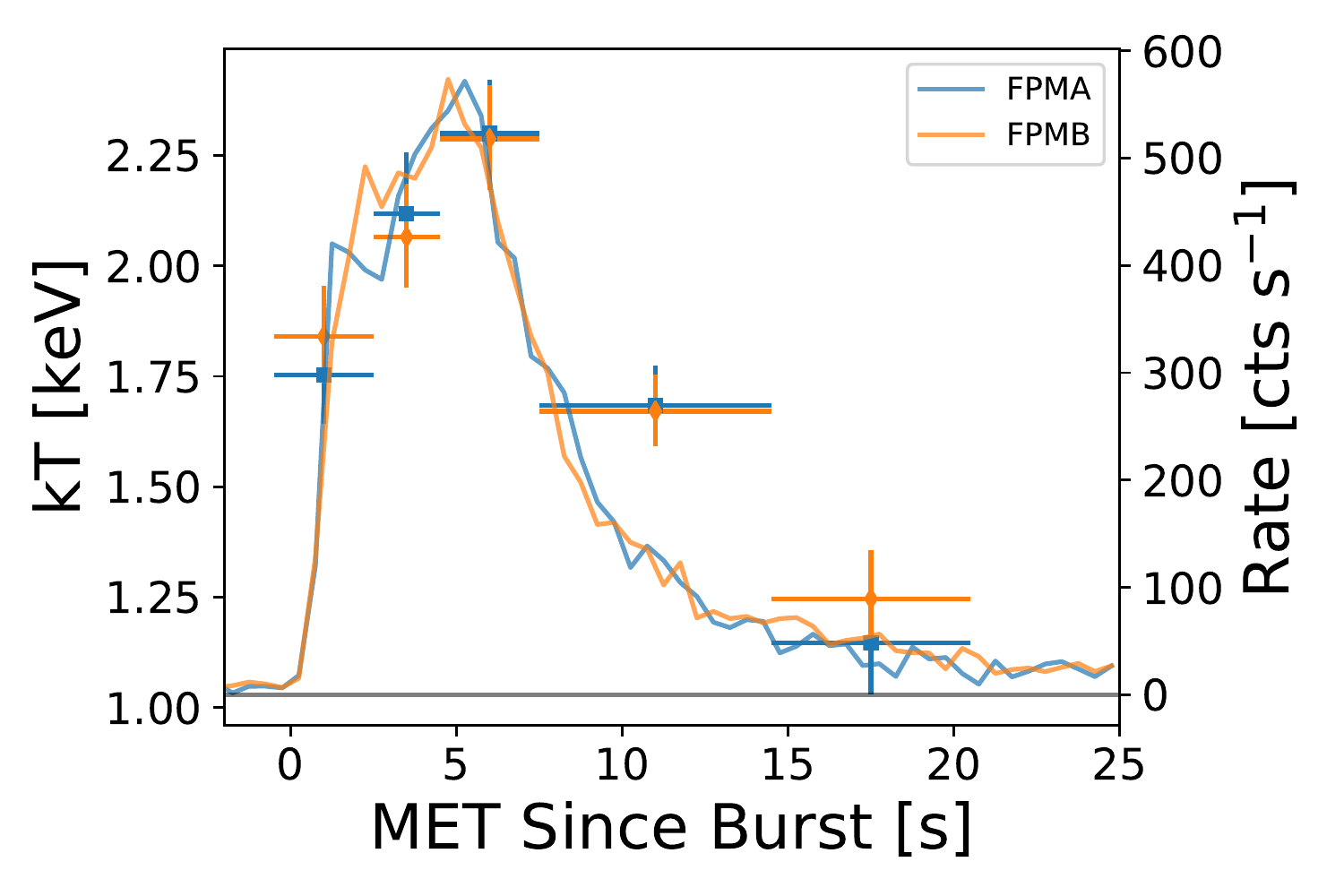} 
    \includegraphics[width=0.32\linewidth]{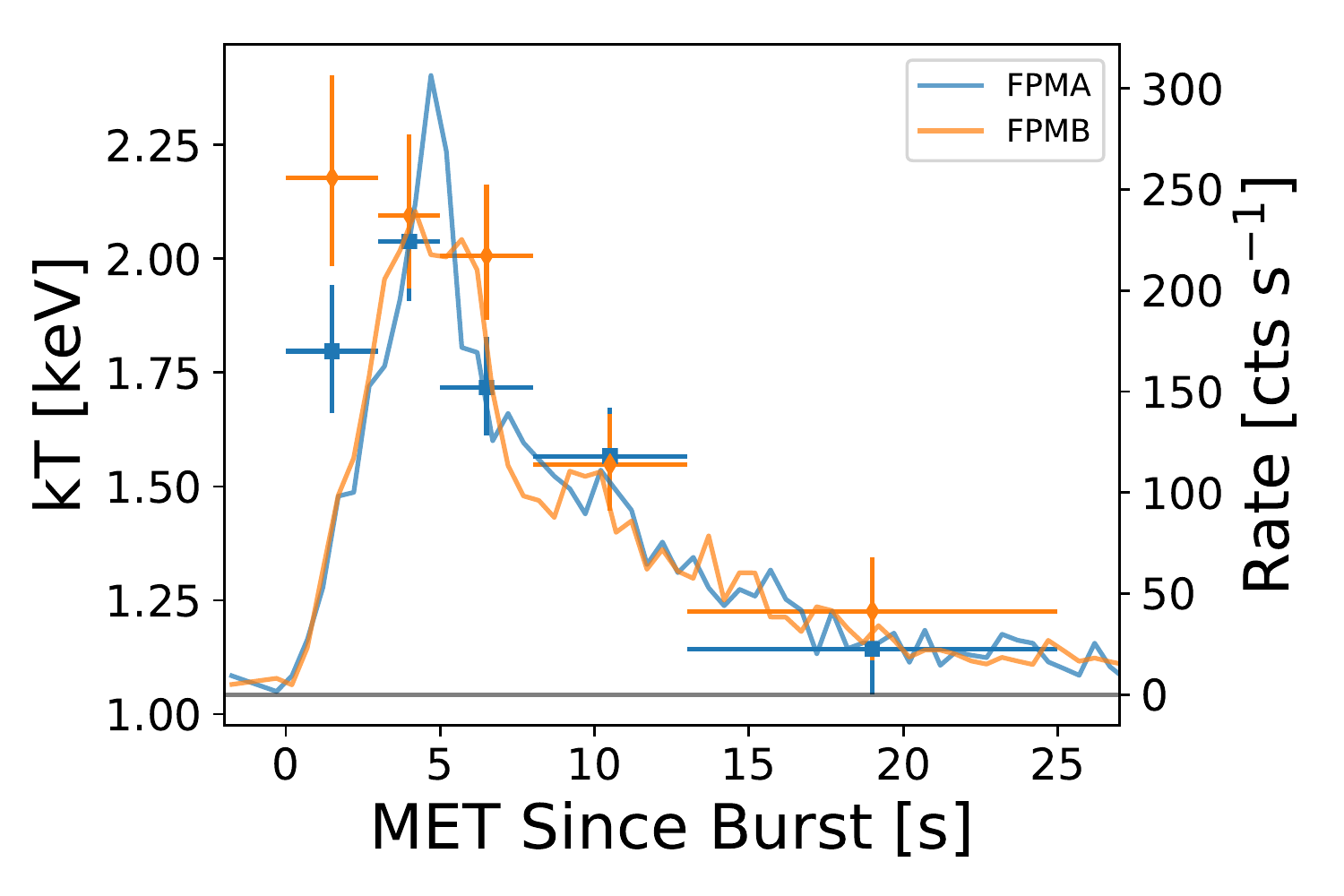} 
    \includegraphics[width=0.32\linewidth]{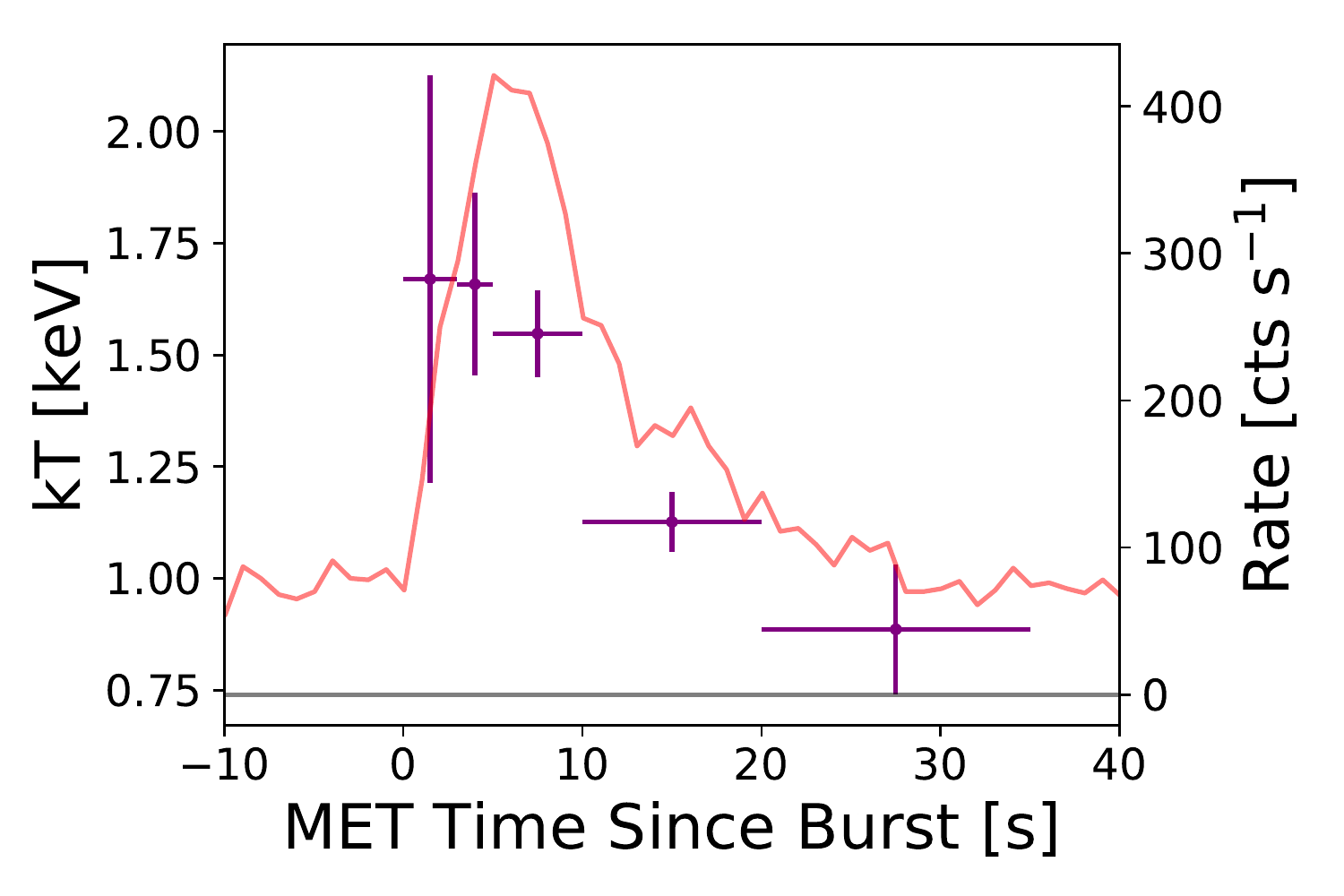} 
	\caption{The BB2 kT parameter for (\textit{Left:}) burst 2 (\textit{Center:}) burst 3 (\textit{Right:}) burst 17. Overplotted are the count rates in 1~s bins, with the corresponding scale on the right side vertical axis.}
	\label{fig: kT evolution of nustar bursts}
\end{figure}

\begin{figure}[!b]
	\centering
    \includegraphics[width=0.46\linewidth]{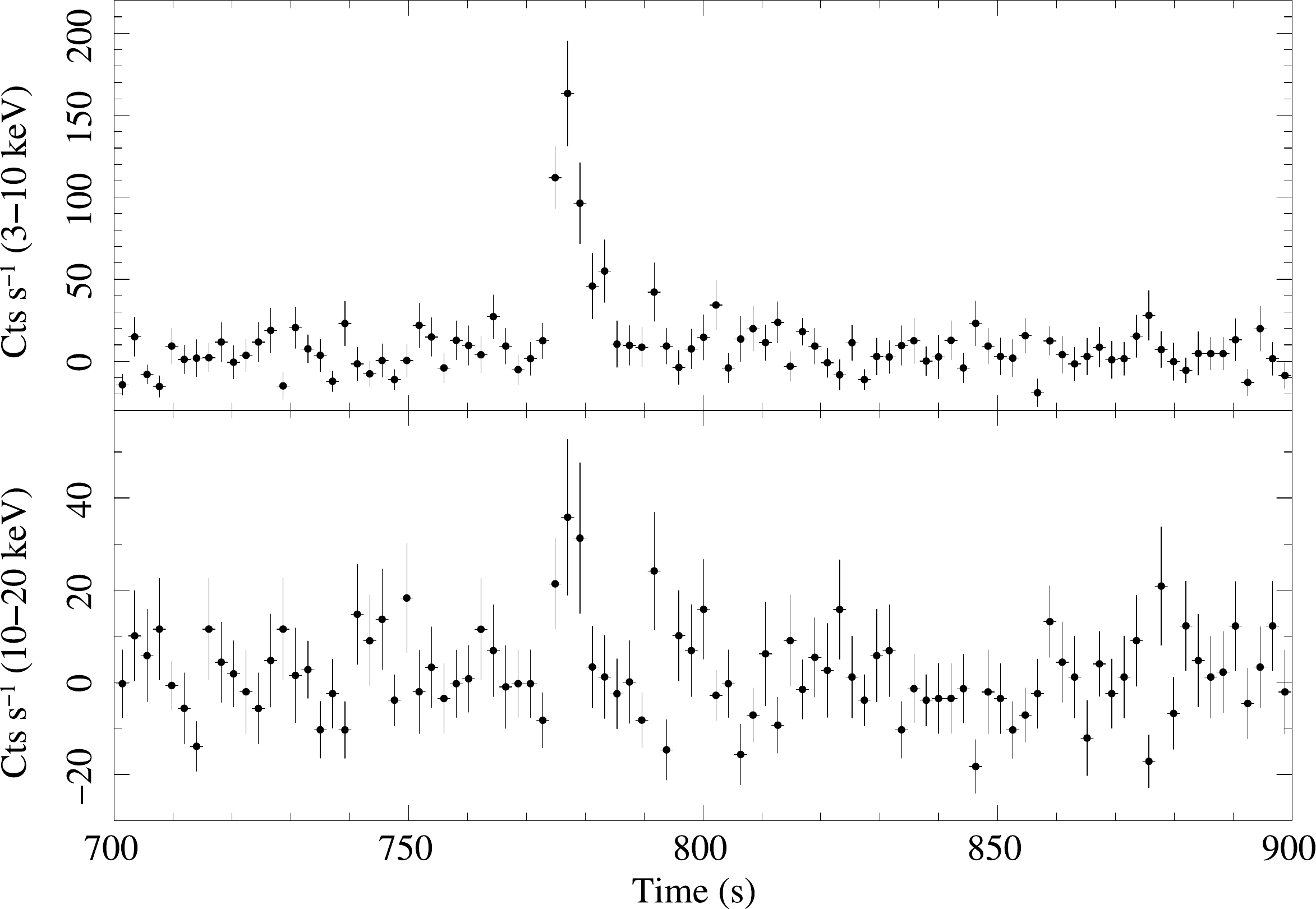}    
    \includegraphics[width=0.46\linewidth]{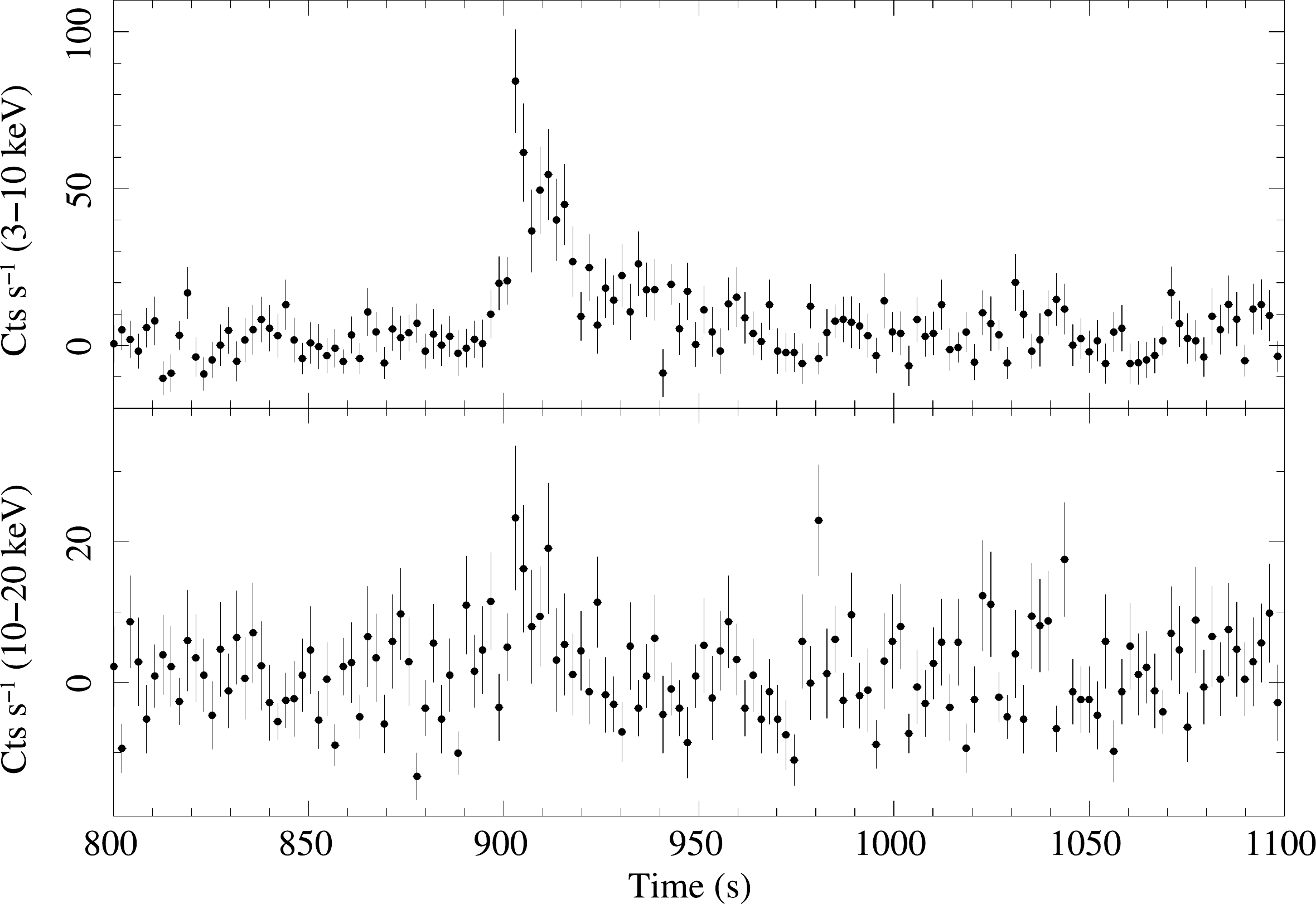}
	\caption{Two examples of bursts observed by JEM-X. We show on the left the 11th burst that achieved the highest peak flux (see text for details). On the right, we show the burst that was reported to have undergone a photospheric radius expansion but for which we concluded that the statistics are too low to draw a firm conclusion.}
	\label{fig:jmx_burst}
\end{figure}



\begin{table}[!t]
\centering
{\normalsize \begin{tabular}{lccc}
Burst    & Instrument       & Onset Time         &   Day\\
         &                  & UTC                &   MJD\\ \hline\hline
         
 1.       &  \maxi{}/Cam2   &        19 Oct. 2017 11:36:52      &  58045.48393\\ 
 2.       &  \nustar{}/FPMA+B  &      03 Dec. 2017 01:29:17      &  58090.06200\\ 
 3.       &  \nustar{}/FPMA+B  &       03 Dec. 2017 05:01:02      &  58090.20905\\
 4.       &  \maxi{}/Cam2   &        24 Dec. 2017 03:35:00      &  58111.14930\\        
 5.       &  \maxi{}/Cam2   &        31 Dec. 2017 14:13:24      &  58118.59263\\
 6.       &  \integral{}     &        30 Jan. 2018 15:19:48      &  58148.63875\\ 
 7.       &  \maxi{}/Cam1+2+7   &       30 Jan. 2018 21:13:00      &  58148.88402\\
          &  \integral{}     &        30 Jan. 2018 21:13:02      &  58148.88405\\
 8.       &  \integral{}     &        31 Jan. 2018 01:39:22      &  58149.06900\\ 
 9.       &  \integral{}     &        03 Feb. 2018 21:51:07      &  58152.91050\\ 
 10.      &  \integral{}     &        06 Feb. 2018 03:42:09      &  58155.15427\\ 
 11.      &  \integral{}     &        06 Feb. 2018 08:01:32      &  58155.33440\\ 
 12.      &  \integral{}     &        08 Feb. 2018 06:30:15      &  58157.27101\\ 
 13.      &  \integral{}     &        08 Feb. 2018 13:37:40      &  58157.56782\\ 
 14.      &  \integral{}     &        08 Feb. 2018 16:40:52      &  58157.69505\\ 
 15.      &  \integral{}     &        09 Feb. 2018 01:29:09      &  58158.06191\\ 
 16.      &  \integral{}     &        10 Feb. 2018 18:20:47      &  58159.76443\\ 
 17.      &  \nicer{}/XTI    &        22 Feb. 2018 22:34:00      &  58171.94247\\ 
 18.      &  \maxi{}/Cam1+2+7 &        18 Apr. 2018 19:42:03      &  58226.82086\\
 19.      &  \maxi{}/Cam2   &        24 May 2018 22:44:23       &  58262.94748\\
 20.      &  \maxi{}/Cam5   &        01 Jun. 2018 11:36:03      &  58270.48336\\
 21.      &  \maxi{}/Cam2   &        05 Aug. 2018 11:36:39      &  58335.48378\\
 22.      &  \maxi{}/Cam5   &        12 Aug. 2018 15:16:16      &  58342.63629\\
 23.      &  \maxi{}/Cam1+2+7    &        02 Sep. 2018 05:02:12      &  58363.20986\\
 24.      &  \maxi{}/Cam2   &        16 Oct. 2018 13:57:02      &  58407.58127  \\\hline\hline

\end{tabular} } 
\caption{A list of the type I X-ray bursts detected from J1621.}
\label{tbl: bursts list}
\end{table}

\section{Type I X-ray Bursts} \label{sec: bursts}

Twenty-four type I X-ray bursts were observed from \maxinos{} by four instruments: 11 with \maxi{}, 11 with \integral{}, 2 with \nustar{}, and 1 with \nicer{}. One burst was seen with both \integral{} and \maxi{}. This unambiguously identifies \maxinos{} as a system hosting a neutron star undergoing nuclear burning on the surface. Table~\ref{tbl: bursts list} exhibits all burst onset times, dates, and the detecting instrument(s).


\subsection{Light Curves}

We measured the duration of two \nustar{} bursts (temporally binned at 1~s) by using the  $T_{\rm 90}$ method, first developed by \citet{kouv93} for Gamma-Ray Burst duration measurements. Burst 2 and 3 (Tbl. \ref{tbl: bursts list}) durations were found to be 19$\pm{}$2~s, and 24$\pm{}$2~s, respectively. We carried out the same approach with the burst observed with \nicer{} and found $T_{90}=$ 33$\pm{}$2~s. We note that these differences in burst duration may come from the spectral range to which each instrument is sensitive ; longer durations are expected  in softer energy bands for the typical BB spectrum that reaches kT$\sim{}$2~keV \citep{Lewin+1993}, which is what we observe.

In the \nustar{} energy range, we expect a similar burst duration distribution to that of \integral{}/JEM-X (3-20~keV) \citet[top left panel of Fig. 4][]{Chelovekov+2017}. The \nustar{} burst durations are longer than most observed with JEM-X, the distribution of which peaks at $\sim{}10$~s. Despite a nonuniform energy range for these comparisons, all three burst durations lie within the range of $10-100$~s, where 111/159 $\approx{}$70\% 	have been recorded \citep[Table 2 of][]{Galloway_Keek2017}.

The \maxi{} observations totaled $\sim{}$144~ks over 300 days, during which we found 11 significant type I bursts, with average peak count rate of $\sim{}$2 cts s$^{-1}$cm$^{-2}$ (2-20~keV band).

\begin{table}[!t]
\begin{tabular}{cccccc}
Bin & $N_{\text{H}}$ &kT$_{BB}$ &F$_{1\text{-}10}$ &$\chi{}^{2}$/dof & red. $\chi{}^{2}$ \\
s &10$^{22}$~cm$^{-2}$& keV & $10^{-9}$ & & \\ \hline\hline
1.0 - 3.0 & 6.1$\pm{}$0.3 & 1.40$^{+0.34}_{-0.23}$ & 1.86$\pm{}$0.28&  228 / 218 & 1.06 \\
3.0 - 5.0 & Linked & 1.45$^{+0.19}_{-0.15}$ & 5.57$\pm{}$0.53&  230 / 224 & 1.04 \\
5.0 - 10.0 & Linked & 1.49$\pm{}$0.09 & 6.97$\pm{}$0.37 & 262 / 261 & 1.01 \\
10.0 - 20.0 & Linked &  1.11$\pm{}$0.06 & 2.33$\pm{}$0.16 & 284 / 260 & 1.09 \\
20.0 - 35.0 & Linked & 0.81$\pm{}$0.07 & 0.48$\pm{}$0.08 & 251 / 243 & 1.03 \\
35.0 - 50.0 & Linked & - & $<$ 0.50 & 246 / 237 & 1.05 \\ \hline\hline
\end{tabular}
\caption{The spectral parameters best fit to six time bins (time since burst onset) within the type I X-ray burst observed with \nicer{}/XTI.}
\label{tbl: NICER burst}
\end{table}

\subsection{Spectroscopy}

We fitted the \nustar{} burst spectra with two components: an absorbed BB plus disk reflection, frozen at the best fit parameters of the persistent emission spectrum (see $\S{}$\ref{section:persistent x-ray spectroscopy}) and a second absorbed BB (BB2). We split each burst into 5 intervals chosen to cover the rise part of the burst (two bins), its peak, and its decay (two bins). The $kT$ evolution of the BB2 component is shown in Fig. \ref{fig: kT evolution of nustar bursts}, left and center panels.  

Burst 17 was observed during obs. 26 with \nicer{}. To observe the full burst, we relaxed the bright Earth elevation filter to those data taken $>$ 35$^{\circ{}}$ from the limb. To establish the persistent level, we fitted 125~s of pre-burst emission, which is well described by a PL with $\Gamma{}=2.70\pm{}0.10$ and $N_{\text{H}}$=(6.1$\pm{}0.3$)\E{22} cm$^{-2} $($\chi{}^{2}$/dof=216/201=1.08). The persistent flux was at F$_{1-10}$=(2.07$\pm{}0.16$)\E{-9}\,erg s$^{-1}$ cm$^{-2}$. Freezing these parameters, we added a BB component and carried out a time-resolved approach for 6 time bins. The result is shown in Tbl. \ref{tbl: NICER burst} and at the rightmost panel of Fig. \ref{fig: kT evolution of nustar bursts}. All three bursts showed a similar $kT$ evolution. 


For the brightest \integral{} burst, we obtained the effective-area corrected peak flux of $1.8 \pm 0.3$~cts~s$^{-1}$~cm$^{-2}$ at 2-10~keV, which corresponds to $F_{bol}\sim$2.3\E{-8}~ergs~s$^{-1}$~cm$^{-2}$. Temporally, the burst showed a 10~s monotonic rise followed by an exponential decay (e$^{-\tau{}/1\text{s}}$, with $\tau{}=9\pm{}3$~s). This gave an effective burst duration of $\tau{}_{b}=18$~s.


In order to search for additional bursts observed with \integral{}, we extracted the JEM-X1 and JEM-X2 lightcurves with a time resolution of 2~s. A total of 11 bursts were found \citep[see also][]{integral_sees_burst}, and we report the onset time of all these events in Table~\ref{tbl: bursts list}. The bursts from the source were relatively faint for JEM-X and we could extract a meaningful spectrum during the 8~s around the peak only for the 11th burst, which was also the brightest (reaching about 150~cts~s$^{-1}$ in the 3-20 keV energy band; note that integrations shorter than 8~s are not possible with the standard OSA software). We fitted the JEM-X1 and JEM-X2 spectra with a BB model (the absorption column density was fixed to $2.5\times10^{22}$~cm$^{-2}$). We used in the fit as a background the spectrum extracted during the remaining available exposure time of the SCW, where the bust was identified (SCW ID. 191800230010). We measured a BB temperature of $kT$=1.9$\pm$0.3~keV, a radius of 13.5$\pm$3.0~km (assuming a distance of 8.4~kpc), and a 3-20~keV flux of  (3.2$\pm$0.6)$\times$10$^{-8}$~erg~cm$^2$~s$^{-1}$ (all uncertainties are given at 90\% c.l.). We did not find evidence of a clear photospheric radius expansion in any of the JEM-X bursts. \citet{integral_sees_burst} mentioned that the burst of 2018 February 3 at 21:51:07 might have undergone a photospheric radius expansion, but we show in Fig. \ref{fig:jmx_burst} that the statistical quality of this event is too low to draw any firm conclusions.

\subsection{X-ray Timing}

We searched the burst 17, seen with \nicer{}, for burst oscillations. For our search we set up a sliding window of length T and stride S = T /2. The number of strides is set such that the last window is at most 35 seconds after the burst onset. For each window we computed the power spectrum and considered the power spectral bins for frequencies between 50 Hz and 1000 Hz. We then compared the obtained powers with a detection threshold treating all trials (counting every spectral bin, of every window stride) as though they were independent \citep[see, e.g.,][for a description of power spectrum detection thresholds]{vanderKlis1989}. We applied this search strategy for T = 2, 4, and 8, but no burst oscillations were detected.

\section{Discussion}\label{discussion}

\maxinos{} was discovered with \maxi{} on 19 October, 2017 at an X-ray flux approximately four orders of magnitude higher than its deepest upper limit emission in quiescence. It is the first DGPS transient which we followed up with a comprehensive multi-wavelength observational campaign to identify its nature. The source was successfully classified as the 111th Type I X-ray burster\footnote{see also the web page of Jean In't Zandt: \href{https://personal.sron.nl/~jeanz/bursterlist.html}{https://personal.sron.nl/\~{}jeanz/bursterlist.html}}, after it was detected to emit type I X-ray bursts, soon after its outburst \citep{Bult2018a}; in the following 15 months, a total of 22 additional bursts were detected with four separate X-ray instruments.

The source persistent emission spectrum can be adequately described with a three component model: an absorbed thermal (BB), a nonthermal (PL), and an emission feature (fit with a Lorentzian centered at $\sim{}$6.4~keV) indicating an ionized Fe reflection line from an accretion disc. There is clear spectral evolution during the outburst, with the hardest spectra appearing at the rising part of the initial outburst ($\Gamma=1.4$), while the remaining available spectra cluster around $\Gamma{}\sim{}2$. We note here, however, that we only had good coverage of the light curve at the beginning of the outburst and sporadic \swift/XRT data thereafter. 

The X-ray light curve of the source appears to be episodic, with at least 6 distinct peaks separated at $\sim78$ days.  Simultaneous \swift{}/XRT and \integral{} observations confirm the episodic nature of the source with one apparent discrepancy: during $\sim$ MJD 58150 and 58159 (see Fig. \ref{fig. lightcurve}), there is a flux rise in the \swift/XRT accompanied with a similar rise in the \integral{} light curve. However, immediately after the peak, the source is not detected with \integral{}, while it is still well detected with XRT. 
We attribute this increase of the non-thermal photon intensity to inverse Compton scattering in a hot free-electron halo surrounding the NS. The incident thermal spectrum from the 11 bursts emitted during this interval would have provided a large photon flux, which was subsequently upscattered to the \integral{}/IBIS energy range on a short timescale, due to the impulsive nature of the burst. After the paucity of bursts, the soft X-rays declined slowly, while the hard X-rays disappeared rapidly.

Besides the long-timescale lightcurve modulation, a noteworthy characteristic of the two \integral{} lightcurves in Figure 3 is the moderate disagreement with that in the low-energy band. This suggests a somewhat different origin for contributions above and below 10 keV.  The spectroscopic fitting of the Swift/XRT and NICER data presents the case at low energies that there is a mix of spectral components below 10 keV.  At higher energies, the power law shape in the \integral{} spectra (see Figure \ref{fig:integralspe}) is much simpler to interpret.  It is perhaps suggestive of inverse Compton emission produced by non-thermal relativistic electrons.  Alternatively, and probably more appropriate for accreting systems that have moderate to high opacities, it resembles the classic unsaturated Comptonization spectrum realized in models of accreting black holes such as in Cyg X-1 or in active galactic nuclei.  The power law arises due to repeated scatterings of lower energy photons by hot, thermal electrons of temperature $T_e$ that slowly increases the photon energy until it is close to $kT_e$. The power law marks the scale-independence of the Compton upscattering, and its slope depends only on the mean energy gain per collision, $\langle \Delta E\rangle =4 kT_e$ for non-relativistic electrons, and the probability of loss of photons from the scattering zone.\footnote{The interested reader may wish to consult Chapter 7 of \citet{Rybicki+Lightman1979} for a summary of its development as a solution of the Kompaneets equation.} The resulting differential photon spectrum is described by
\begin{equation}
    \frac{dN}{dE}\;\propto\; E^{-\alpha}
    \quad ,\quad
    \alpha\; =\; -\frac{1}{2} + \sqrt{\frac{9}{4} + \frac{4}{y}}
    \quad ,\quad
    y \; =\; \frac{4 kT_e}{m_ec^2}\, \hbox{max} \{\tau,\, \tau^2\} \;\; ,
  \label{Comp_index}
\end{equation}
with the Compton $y$-parameter normally in the domain $y< 1$. This parameter is the product of the average fractional energy change per scattering and the mean number of Thomson scatterings, and $\tau$ is the scattering Thomson optical depth. The index $\alpha$ is a declining function of $y$. The extension of the power law persists until an exponential turnover arises at $E\sim kT_e$.  

If such a coronal Comptonization picture is used to interpret the \integral{} spectra, then the index provides a measure of the opacity and/or the temperature. The measured value of $\alpha\sim 1.3\pm 0.5$ during revolution 1914-1917 suggests a value $y\sim 4-5$. Temporally, one expects coronae proximate to an accretion disk to be quite variable, perhaps due to magnetic field line flaring activity, much like the solar corona with its mass ejections.  The field can be a source for energization of the system.  The result is varying or chaotic time profiles.  This is consistent with the \integral{} ISGRI fluxes presented in Figure 3. Flux variations probably trace coronal electron heating rates since the seed photons of disk origin should be approximately constant in luminosity. Enhanced fluxes produced by electron density $n_e$ increases would raise $\tau$, trapping photons more effectively in the Comptonizing cloud and hardening the emergent spectra (lower $\alpha$).  Similar character would be realized by hotter electrons. This degeneracy of information can only be disentangled with the observation of a spectral turnover at different epochs, thereby constraining $T_e$ as a function of time. Unfortunately, the \integral{} spectra do not clearly exhibit such quasi-exponential turnovers, so that $kT_e \gtrsim 100\,$keV is inferred.  

Volumetric influences complicate this picture somewhat.
It is quite possible that magnetic squeezing of electrons by mobile
field lines can adiabatically increase the density and temperature of
the hot electrons simultaneously. A noteworthy characteristic of the two \integral{} lightcurves in Figure 3 is that the hardness ratio (and therefore $\alpha$) does not in fact vary much with time. Then the flux
variability and implied spectral constancy could be driven by density
fluctuations coupled to changes in the effective volume $V$ of the
Comptonization zone. With $\tau \propto n_e V^{1/3}$ ($\lesssim 1$), if
$n_e \propto V^{-\beta}$, then one infers $T_e \propto V^{\beta - 1/3}$
in order to keep the Compton $y$ parameter approximately constant.  For
plasma flow connected to divergent/convergent coronal field lines, values of
$\beta \sim 2/3$ are expected for wind-like expansions/contractions,
indicating that small or modest temperature changes should accompany the
observed variability.  In particular, volume contractions should induce
coupled increases in both density $n_e$ and temperature $T_e$, with
$T_e\propto (n_e)^{1/2}$.  This coupling defines a potential diagnostic of the coronal interpretation, though to bring it to fruition requires a more sensitive hard X-ray/soft gamma-ray telescope.


The episodic nature of the observed outbursts is intriguing. A possible explanation for the 78-day variations in its light curve may be the so-called `super-orbital periods' or long periods. These have been noted in a number of low- and high-mass X-ray binaries. A better name for them would be `long time scale modulations', since very often they are not strictly periodic; individual modulations in the \maxinos{} lightcurve vary from approximately 50 to 90 days in duration. For quite a few systems there is a broad correlation of this long timescale modulation with orbital period, though with a fair amount of scatter \citep{Sood+2007}. The ratio of long timescale to orbital period ranges from 10--100 in these systems (WP99). In some cases the ratio is much greater, e.g.\ in 4U 1820--30, where the long timescale is 176\,d, for an orbital period of 11\,min (ratio 23,000). WP99 demonstrate that these periods can be  explained reasonably well by a combination of disk irradiation by the central source, causing it to tilt and warp, and tidal torque from the companion, further driving the precession of this tilted disk.

To test whether the long time scale here would fit the radiative-precession model, we can use eqs.\ 17--19 of WP99, provided we know the properties of \maxinos{} well enough. From the IR data, the orbital period is estimated to be in the range 3--20~h (Bahramian et al., in prep), implying that the companion is low-mass and on or just beyond the main sequence, and we thus infer a companion mass in the range 0.3--1~M$_\odot$. The accretor is a neutron star, for which we assume a mass of 1.4~M$_\odot$. The X-ray luminosity, assuming an upper limit of the distance of 5~kpc derived from IR data (Bahramian et al., in prep), is in the range  $0.45-5.98\times10^{36}$~erg s$^{-1}$.  Making the same assumptions as WP99 for the outer disk radius, we can compute the radiative precession period of the disk in this system to be

\begin{equation}
 P_\Gamma = 82\,{\rm days}\;\; \alpha_{-1}^{-4/5} 
              \left(\frac{\epsilon}{0.2}\right)^{-1}
              L_{X,36.5}^{-0.3}
              \left(\frac{P_{\rm orb}}{12\,\rm h}\right)^{2/3}
              \left(\frac{M_T}{2\,M_\odot}\right)^{1/3}  
\end{equation}
  
Here numerical subindices indicate logarithms of normalization values. We have chosen standard neutron star values $M=1.4~M_\odot$, $R=10$~km, corresponding to an accretion efficiency $\epsilon=0.2$ to convert between X-ray luminosity and mass accretion rate, and normalized to middle-of-range values for the X-ray luminosity, orbital period, and total mass of the system. We see that for reasonable values of the system parameters, the 82~day radiative precession period we predict is close to the observed long time scale modulation of 78~d. This result supports a super-orbital period as the underlying model for the observed lightcurve modulation.

\appendix
\section{Spectral Simulations Plots}
\begin{figure}[!h]
	\centering
	\includegraphics[scale=0.8]{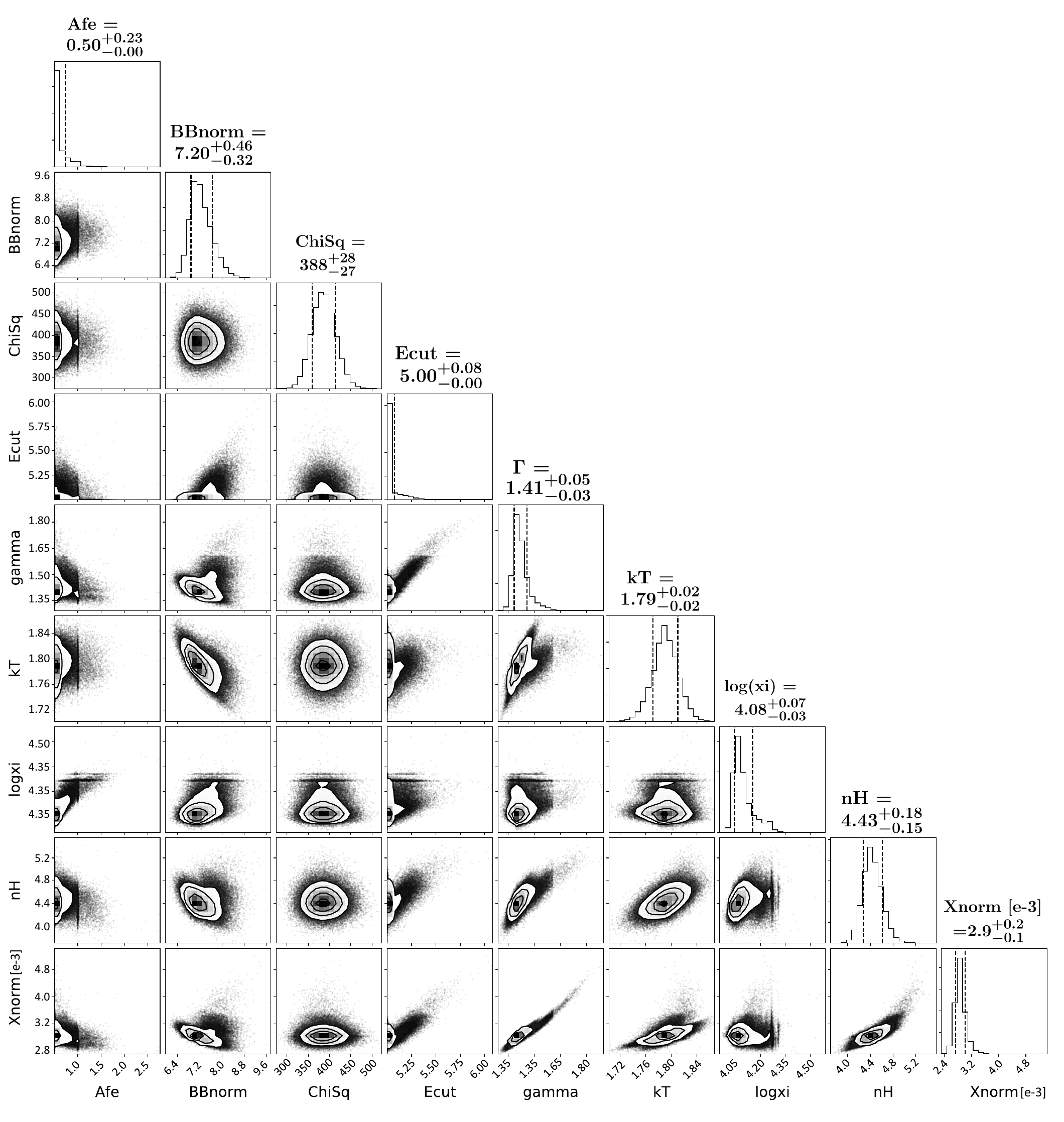}
	\caption{Parameters for 10$^{5}$ simulated spectra using an absorbed reflection spectrum plus blackbody model. Contours are at the volumetric 1$\sigma{}$ level, 19.7\% to either side of the centroid.}
	\label{fig: dim_xillver_corner}
\end{figure}

\begin{figure}[!h]
	\centering
	\includegraphics[scale=0.8]{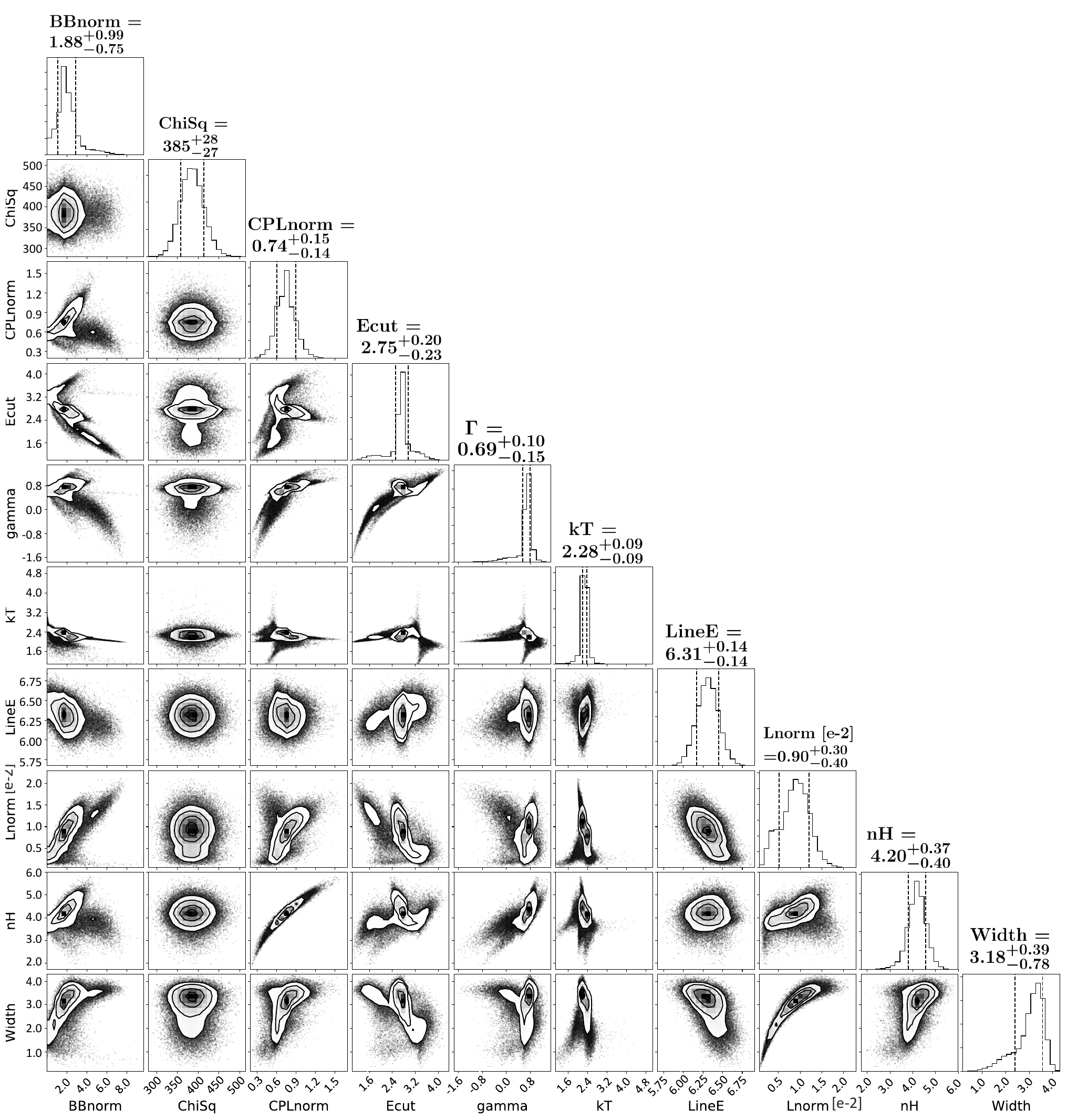}
	\caption{Parameters for 10$^{5}$ simulated spectra using an absorbed cutoff PL plus BB plus a Lorentzian. Contours are at the volumetric 1$\sigma{}$ level, 19.7\% to either side of the centroid.}
	\label{fig: bright_pl+bb+lorentz_corner}
\end{figure}

\acknowledgements{}
\begin{center}
ACKNOWLEDGEMENTS
\end{center}

\noindent
We thank the anonymous referee, whose comments improved the quality of this work. N.G., C.K., and P.B. acknowledge support from  Chandra award number GO8-19061X and Swift award number NNH16ZDA001N-SWIFT. \,The IRSF project is supported by the Grants-in-Aid for Scientific Research on Priority Areas (A) (No. 10147207 and No. 10147214).
\;H.N. acknowledges support from a Grant-in-Aid for Scientific Research (No. 16K05301).
\;The Australia Telescope Compact Array is part of the Australia Telescope National Facility which is funded by the Australian Government for operation as a National Facility managed by CSIRO.
\;Special thanks go to Nobuyuki Kawai, Christina Gilligan, and Jill Neeley for their indispensable roles in data acquisition for this project. D.H. acknowledges support from the DIRAC Institute in the Department of Astronomy at the University of Washington. The DIRAC Institute is supported through generous gifts from the Charles and Lisa Simonyi Fund for Arts and Sciences, and the Washington Research Foundation. This work made use of data from the {\it NuSTAR} mission, a project led by the California Institute of Technology, managed by the Jet Propulsion Laboratory, and funded by the National Aeronautics and Space Administration. This research has made use of the {\it NuSTAR}  Data Analysis Software (NuSTARDAS) jointly developed by the ASI Science Data Center (ASDC, Italy) and the California Institute of Technology (USA). We thank all operations, software and calibration teams contributing to the facilities used in this study for support with the execution and analysis of these observations.


\facilities{\swift{}, \cxo{}, \nustar{}, \nicer{}, \maxi{}, \integral{}, \atca{}, \irsf{}, \gemini{}, ADS, HEASARC}
\software{MIRIAD (Sault et al. 1995), Stingray (Huppenkothen et al. 2016, 2019), emcee (Foreman-Mackey et al. 2013), CIAO (v4.9.3; Fruscione et al. 2006), HEAsoft (v624; HEASARC 2014), XSPEC (v12.10.0; Arnaud 1996)}


\end{document}